\documentclass[10pt]{article}
\usepackage{graphicx}
\usepackage{amsmath}
\usepackage{amssymb}
\usepackage{caption2}
\begin{document}
\begin{center}
{\large {\bf \sc{  Analysis of the strong decays $D_{s3}^*(2860)\to DK$, $D^{*}K$   with QCD sum rules }}} \\[2mm]
Zhi-Gang Wang \footnote{E-mail,zgwang@aliyun.com.  }      \\
 Department of Physics, North China Electric Power University,
Baoding 071003, P. R. China
\end{center}

\begin{abstract}
In this article, we assign the $D_{s3}^*(2860)$ to be a D-wave $c\bar{s}$  meson, study  the  hadronic coupling constants $G_{D_{s3}^*(2860)DK}$ and $G_{D_{s3}^*(2860)D^*K}$   with the three-point QCD sum rules, and calculate  the partial  decay widths $\Gamma\left(D_{s3}^*(2860)\to D^{*}K\right)$ and $\Gamma\left(D_{s3}^*(2860)\to DK\right)$. The predicted  ratio $R=\Gamma\left(D_{s3}^*(2860)\to D^{*}K\right)/\Gamma\left(D_{s3}^*(2860)\to DK\right)=0.57\pm0.38$  cannot reproduce the experimental value  $R={\rm Br}\left(D_{sJ}^*(2860)\to D^{*}K\right)/{\rm Br}\left(D_{sJ}^*(2860)\to DK\right)=1.10 \pm 0.15 \pm 0.19$.
\end{abstract}



\section{Introduction}

In 2006, the BaBar  collaboration  observed the $D^*_{sJ}(2860)$ meson in decays  to the final states  $D^0 K^+$ and $D^+K^0_S$, the measured  mass and width are $(2856.6 \pm 1.5  \pm 5.0)\, \rm{ MeV}$ and $(48 \pm 7 \pm 10)\, \rm{ MeV}$, respectively \cite{BaBar2006}.
In 2009, the BaBar  collaboration confirmed the $D^*_{sJ}(2860)$  in the $D^*K$ channel, and measured the ratio $R$ among the branching fractions \cite{BaBar2009},
 \begin{eqnarray}
R&=& \frac{{\rm Br}\left(D_{sJ}^*(2860)\to D^*K\right)}{{\rm Br}\left(D_{sJ}^*(2860)\to D K\right)}=1.10 \pm 0.15 \pm 0.19\, \, .
 \end{eqnarray}
The observation of the  decays $D^*_{sJ}(2860)\to D^*K$ rules out the $J^P=0^+$ assignment, the possible assignments are
   the $1^3{\rm D}_3$ $c\bar{s}$ meson \cite{Colangelo0607,Zhong2008,Zhong2010,Zhang2007,Li2007,Chen2009,Li0911,Badalian2011}, the $c\bar{s}-cn\bar{s}\bar{n}$ mixing state \cite{Vijande2009}, the dynamically generated  $D_1(2420)K$ bound state \cite{FKGuo}, etc.

In 2014,  the LHCb collaboration  observed a structure at $2.86\,\rm{GeV}$  in the $\overline{D}^0K^-$ mass distribution  in the Dalitz plot analysis of the decays $B_s^0\to \overline{D}^0K^-\pi^+$,  the structure contains both spin-1 (the $D_{s1}^{*-}(2860)$) and spin-3 (the $D_{s3}^{*-}(2860)$) components \cite{LHCb7574,LHCb7712}.
Furthermore, the LHCb collaboration obtained the conclusion that the $D^*_{sJ}(2860)$ observed  in the inclusive $e^+e^- \to \overline{D}^0K^{-}X$ production by the BaBar collaboration and  in the $pp \to \overline{D}^0K^{-}X$ processes by the LHCb collaboration consists of at least two particles  \cite{BaBar2009,LHCb1207}.

The  QCD sum rules is a powerful  theoretical tool in studying the
ground state hadrons and has given many successful descriptions of the masses, decay constants, form-factors and hadronic coupling constants, etc \cite{SVZ79,Reinders85}.
In Ref.\cite{Wang1603}, we assign the $D_{s3}^*(2860)$ to be a D-wave $c\bar{s}$  meson, and  study the mass and decay constant (or the current-meson coupling constant) of the $D_{s3}^*(2860)$ with the QCD sum rules. The predicted mass $M_{D_{s3}^*}=(2.86\pm0.10)\,\rm{GeV}$ is in excellent agreement with the experimental value  $M_{D_{s3}^*}=(2860.5\pm 2.6 \pm 2.5\pm 6.0)\,\rm{ MeV}$ from the LHCb collaboration \cite{LHCb7574,LHCb7712}. We obtain further support by reproducing the mass of the $D_{s3}^*(2860)$ based on the QCD sum rules.

If we assign the $D^*_{sJ}(2860)$ to be the $1^3{\rm D}_3$ state, the ratio $R$ from the leading order
 heavy meson effective theory \cite{Colangelo0607}, the constituent quark model with quark-meson effective Lagrangians \cite{Zhong2010},
 the ${}^3{\rm P}_0$ model \cite{Zhang2007,Li0911,Song2014,Segovia1502,Chen1507} and the relativized quark model  \cite{Godfrey2013} cannot reproduce the experimental value $R=1.10 \pm 0.15 \pm 0.19$ \cite{BaBar2009}. The values of the ratio $R$ from different theoretical methods are shown explicitly in Table 1. From the table, we can see that even in the ${}^3{\rm P}_0$ model the predictions are quite different, as different harmonic oscillator wave-functions are chosen to approximate the mesons' wave-functions.

\begin{table}
\begin{center}
\begin{tabular}{|c|c|c|c|}\hline\hline
       R                        & Theoretical methods $\&$ experimental data \\   \hline
   $1.10 \pm 0.15 \pm 0.19$     & Experimental value from BaBar  \cite{BaBar2009}         \\  \hline
   $0.39$                       & Leading order heavy meson effective theory \cite{Colangelo0607}  \\ \hline
   $0.40$                       & Constituent quark model with  effective Lagrangians \cite{Zhong2010}\\ \hline
   $0.59$                       & ${}^3{\rm P}_0$ model \cite{Zhang2007}     \\ \hline
   $0.75$                       & ${}^3{\rm P}_0$ model \cite{Li0911}     \\ \hline
   $0.55-0.80$                  & ${}^3{\rm P}_0$ model \cite{Song2014}     \\ \hline
   $0.68$                       & ${}^3{\rm P}_0$ model \cite{Segovia1502}     \\ \hline
   $0.43$                       & ${}^3{\rm P}_0$ model \cite{Chen1507}     \\ \hline
   $0.43$                       & Pseudoscalar emission decay model  \cite{Godfrey2013}   \\ \hline
   $1.10 \pm 0.15 \pm 0.19$     & Heavy meson  effective theory with  chiral symmetry breaking corrections  \cite{WangEPJC2860}  \\   \hline \hline
\end{tabular}
\end{center}
\caption{ The  values of the ratio $R$ from different theoretical methods compared to the experimental data. }
\end{table}

The  $c{\bar q}$ mesons  can be  sorted in doublets according to  the total
angular momentum of the light antiquark ${\vec s}_\ell$,
${\vec s}_\ell= {\vec s}_{\bar q}+{\vec L} $,  in the heavy quark limit, where the ${\vec
s}_{\bar q}$ and ${\vec L}$ are the light antiquark's  spin and orbital angular momentum, respectively \cite{RevWise}.
  For the D-wave mesons, the  doublets $(D^*_{s1},D_{s2})$ and $(D^{\prime}_{s2},D_{s3}^{ *})$ have the spin-parity
$J^P_{s_\ell}=(1^-,2^-)_{\frac{3}{2}}$ and $(2^-,3^-)_{\frac{5}{2}}$,  respectively.
The following two-body strong decays can take place,
\begin{eqnarray}
D_{s3}^{ *+}&\to&D^{*+}K^0\, , \,\,\,D^{*0}K^+\, , \,\,\,D^{*+}_s \eta\, ,\,\,\,D^{+}K^0\, , \,\,\,D^{0}K^+\, , \,\,\,D^{+}_s \eta\, , \nonumber\\
D_{s2}^{ +}&\to&D^{*+}K^0\, , \,\,\,D^{*0}K^+\, , \,\,\,D^{*+}_s \eta\, , \nonumber\\
D_{s2}^{ \prime +}&\to&D^{*+}K^0\, , \,\,\,D^{*0}K^+\, , \,\,\,D^{*+}_s \eta\, , \nonumber\\
D_{s1}^{ *+}&\to&D^{*+}K^0\, , \,\,\,D^{*0}K^+\, , \,\,\,D^{*+}_s \eta\, ,\,\,\,D^{+}K^0\, , \,\,\,D^{0}K^+\, , \,\,\,D^{+}_s \eta\, .
\end{eqnarray}
In Ref.\cite{WangEPJC2860}, we assign  the $D_{s3}^*(2860)$ and $D_{s1}^*(2860)$ to be the $1^3{\rm D}_3$ and $1^3{\rm D}_1$ $c\bar{s}$ states, respectively,  study   the   strong decays   with the heavy meson  effective theory by taking into account  the chiral symmetry breaking corrections. We can reproduce the experimental value   $R =1.10 \pm 0.15 \pm 0.19$ with suitable hadronic coupling constants $\bar{k}^5_Y$ and $\bar{k}^5_X$,  which describe the chiral symmetry breaking corrections. The coupling constant $\bar{k}^5_X$ in the assignment  $D^*_{sJ}(2860)=D^*_{s1}(2860)$ is much larger  than the coupling constant $\bar{k}^5_Y$ in the assignment $D^*_{sJ}(2860)=D^*_{s3}(2860)$. Naively, we expect smaller chiral symmetry breaking corrections,  the  assignment  $D^*_{sJ}(2860)=D^*_{s3}(2860)$ is preferred \cite{WangEPJC2860}.  On the other hand, if the chiral symmetry breaking effects are small enough to be neglected, we have to include some $D_{s2}^{ +}(2860)$ and $D_{s2}^{ \prime+}(2860)$ components, as  they can only decay to the final states $D^*K$, which can enhance the ratio $R$ efficiently.

 In the article, we take the mass and decay constant (or the current-meson coupling constant) of the $D_{s3}^*(2860)$ from the QCD sum rules as input parameters  \cite{Wang1603}, analyze  the vertices $D_{s3}^*(2860)DK$ and $D_{s3}^*(2860)D^*K$ in details to select  the pertinent  tensor structures, and study the  hadronic coupling constants $G_{D_{s3}^*(2860)DK}$ and  $G_{D_{s3}^*(2860)D^*K}$   with the three-point QCD sum rules.  Then we use the   $G_{D_{s3}^*(2860)DK}$ and  $G_{D_{s3}^*(2860)D^*K}$ to calculate the partial decay widths $\Gamma\left(D_{s3}^*(2860)\to D^{*}K\right)$ and $\Gamma\left(D_{s3}^*(2860)\to DK\right)$ and obtain the ratio $R=\Gamma\left(D_{s3}^*(2860)\to D^{*}K\right)/\Gamma\left(D_{s3}^*(2860)\to DK\right)$, and try to reproduce  the experimental value $R=1.10 \pm 0.15 \pm 0.19$ based on the QCD sum rules so as to obtain additional support for assigning the $D^*_{sJ}(2860)$ to be the $D^*_{s3}(2860)$ \cite{WangEPJC2860}.

The article is arranged as follows:  we derive the QCD sum rules for
the hadronic coupling constants  $G_{D_{s3}^*(2860)DK}$ and  $G_{D_{s3}^*(2860)D^*K}$  in Sect.2;
in Sect.3, we present the numerical results and discussions; and Sect.4 is reserved for our
conclusions.

\section{QCD sum rules for  the hadronic coupling constants  $G_{D_{s3}^*(2860)DK}$ and  $G_{D_{s3}^*(2860)D^*K}$ }

In the following, we write down  the three-point correlation functions
$\Pi_{\mu\nu\rho}(p,p^\prime)$ and $\Pi_{\sigma\mu\nu\rho}(p,p^\prime)$  in the QCD sum rules,
\begin{eqnarray}
\Pi_{\mu\nu\rho}(p,p^\prime)&=&i^2\int d^4xd^4y e^{ip^\prime \cdot x}e^{i(p-p^\prime) \cdot (y-z)} \langle
0|T\left\{J_{5}(x)J_{K}(y)J_{\mu\nu\rho}^{\dagger}(z)\right\}|0\rangle\mid_{z=0} \, , \\
\Pi_{\sigma\mu\nu\rho}(p,p^\prime)&=&i^2\int d^4xd^4y e^{ip^\prime \cdot x}e^{i(p-p^\prime) \cdot (y-z)} \langle
0|T\left\{J_{\sigma}(x)J_{K}(y)J_{\mu\nu\rho}^{\dagger}(z)\right\}|0\rangle\mid_{z=0} \, , \\
J_{ 5}(x)&=& \overline{c}(x)i\gamma_5d(x) \, , \nonumber\\
J_{ \sigma}(x)&=& \overline{c}(x)\gamma_{\sigma} d(x) \, , \nonumber\\
J_{ K}(y)&=& \overline{d}(y)i\gamma_5s(y) \, , \nonumber\\
J_{\mu\nu\rho}(z)&=&\overline{c}(z)\left( \gamma_\mu\stackrel{\leftrightarrow}{D}_\nu\stackrel{\leftrightarrow}{D}_\rho
+\gamma_\nu\stackrel{\leftrightarrow}{D}_\rho\stackrel{\leftrightarrow}{D}_\mu+\gamma_\rho\stackrel{\leftrightarrow}{D}_\mu\stackrel{\leftrightarrow}{D}_\nu \right) s(z) \, , \nonumber
\end{eqnarray}
where  the  currents $J_{ 5}(x)$, $J_{ \sigma}(x)$, $J_{K}(y)$ and $J_{\mu\nu\rho}(z)$  interpolate the  mesons $D$, $D^*$, $K$ and   $D_{s3}^*(2860)$, respectively, $\stackrel{\leftrightarrow}{D}_\mu=\stackrel{\rightarrow}{\partial}_\mu-ig_sG_\mu-\stackrel{\leftarrow}{\partial}_\mu-ig_sG_\mu $, the $G_\mu$ is the gluon field.

The current $J_{\mu\nu\rho}(0)$ has negative parity, and  couples potentially to the $J^P={3}^-$  $\bar{c}s$ meson $D_{s3}^*(2860)$.
Furthermore, the current $J_{\mu\nu\rho}(0)$  also couples potentially to the $J^P={2}^+$, $1^-$, $0^+$   $\bar{c}s$ mesons.
The  current-meson coupling constants or the decay constants  $f_{D^*_{s3}}$, $f_{D^*_{s2}}$, $f_{D^*_{s1}}$ and $f_{D^*_{s0}}$ are defined by
\begin{eqnarray}
 \langle 0|J_{\mu\nu\rho}(0)|D_{s3}^*(p)\rangle&=&f_{D_{s3}^*}\varepsilon_{\mu\nu\rho}(p,s)  \, , \\
\langle 0| J_{\mu\nu\rho}(0)|D_{s2}^*(p)\rangle &=&f_{D_{s2}^*} \left[ p_\mu \varepsilon_{\nu\rho}(p,s)+p_\nu\varepsilon_{\rho\mu}(p,s)+p_\rho\varepsilon_{\mu\nu}(p,s)\right] \, , \nonumber\\
\langle 0| J_{\mu\nu\rho}(0)|D_{s1}^*(p)\rangle &=&f_{D_{s1}^*} \left[ p_\mu p_\nu \varepsilon_{\rho}(p,s)+p_\nu p_\rho \varepsilon_{\mu}(p,s)+p_\rho p_\mu\varepsilon_{\nu}(p,s)\right] \, , \nonumber\\
\langle 0| J_{\mu\nu\rho}(0)|D_{s0}^*(p)\rangle &=&f_{D_{s0}^*}  p_\mu p_\nu p_{\rho} \, ,
\end{eqnarray}
where the $\varepsilon_{\mu\nu\rho}(p,s)$, $\varepsilon_{\mu\nu}(p,s)$ and $\varepsilon_{\mu}(p,s)$ are the mesons' polarization vectors with the following properties \cite{JJZhu},
\begin{eqnarray}
{\rm P}_{\mu\nu\rho\alpha\beta\sigma}&=&\sum_{s} \varepsilon^*_{\mu\nu\rho}(p,s)\varepsilon_{\alpha\beta\sigma}(p,s)  \nonumber\\
&=&\frac{1}{6}\left(\widetilde{g}_{\mu\alpha}\widetilde{g}_{\nu\beta}\widetilde{g}_{\rho\sigma}+\widetilde{g}_{\mu\alpha}\widetilde{g}_{\nu\sigma}\widetilde{g}_{\rho\beta}
+\widetilde{g}_{\mu\beta}\widetilde{g}_{\nu\alpha}\widetilde{g}_{\rho\sigma}   +\widetilde{g}_{\mu\beta}\widetilde{g}_{\nu\sigma}\widetilde{g}_{\rho\alpha}
 +\widetilde{g}_{\mu\sigma}\widetilde{g}_{\nu\alpha}\widetilde{g}_{\rho\beta}+\widetilde{g}_{\mu\sigma}\widetilde{g}_{\nu\beta}\widetilde{g}_{\rho\alpha}\right)\nonumber\\
&&-\frac{1}{15}\left(\widetilde{g}_{\mu\alpha}\widetilde{g}_{\nu\rho}\widetilde{g}_{\beta\sigma}+\widetilde{g}_{\mu\beta}\widetilde{g}_{\nu\rho}\widetilde{g}_{\alpha\sigma}
+\widetilde{g}_{\mu\sigma}\widetilde{g}_{\nu\rho}\widetilde{g}_{\alpha\beta}  +\widetilde{g}_{\nu\alpha}\widetilde{g}_{\mu\rho}\widetilde{g}_{\beta\sigma}
 +\widetilde{g}_{\nu\beta}\widetilde{g}_{\mu\rho}\widetilde{g}_{\alpha\sigma}   +\widetilde{g}_{\nu\sigma}\widetilde{g}_{\mu\rho}\widetilde{g}_{\alpha\beta}\right. \nonumber\\
&&\left. +\widetilde{g}_{\rho\alpha}\widetilde{g}_{\mu\nu}\widetilde{g}_{\beta\sigma}  +\widetilde{g}_{\rho\beta}\widetilde{g}_{\mu\nu}\widetilde{g}_{\alpha\sigma}
         +\widetilde{g}_{\rho\sigma}\widetilde{g}_{\mu\nu}\widetilde{g}_{\alpha\beta}\right) \, , \\
 {\rm P}_{\mu\nu\alpha\beta}&=& \sum_s \varepsilon^*_{\mu\nu}(p,s)\varepsilon_{\alpha\beta}(p,s)=\frac{\widetilde{g}_{\mu\alpha}\widetilde{g}_{\nu\beta}
 +\widetilde{g}_{\mu\beta}\widetilde{g}_{\nu\alpha}}{2}-\frac{\widetilde{g}_{\mu\nu}\widetilde{g}_{\alpha\beta}}{3}  \, , \\
 \widetilde{g}_{\mu\nu} &=&\sum_s \varepsilon^*_{\mu}(p,s)\varepsilon_{\nu}(p,s)=-g_{\mu\nu}+\frac{p_\mu p_\nu}{p^2}  \, .
  \end{eqnarray}

At the phenomenological side, we insert  a complete set of intermediate hadronic states with
the same quantum numbers as the current operators $J_{ 5}(x)$, $J_{ \sigma}(x)$, $J_{K}(y)$ and $J_{\mu\nu\rho}(z)$   into the
correlation functions $\Pi_{\mu\nu\rho}(p,p^\prime)$ and $\Pi_{\sigma\mu\nu\rho}(p,p^\prime)$  to obtain the hadronic representation
\cite{SVZ79,Reinders85}. We isolate all the ground state contributions and write them down  explicitly,
\begin{eqnarray}
\Pi_{\mu\nu\rho}(p,p^\prime)&=&\frac{f_D M_D^2 f_K M_K^2f_{D^*_{s3}}\,\,G_{D^*_{s3}DK}(q^2)}{(m_c+m_d)(m_d+m_s)\left(M_{D}^2-p^{\prime2}\right)\left(M_{K}^2-q^{2}\right)\left(M_{D^*_{s3}}^2-p^2\right)}
 \nonumber\\
&&\left\{ \frac{\left[\lambda\left(M_{D^*_{s3}}^2,M_{D}^2,q^2\right)+10M_{D^*_{s3}}^2M_{D}^2\right]\left( M_{D^*_{s3}}^2+M_{D}^2-q^2\right)}{20M_{D^*_{s3}}^6}\,p_{\mu}p_{\nu}p_{\rho}  \right.\nonumber\\
&&+\frac{\lambda\left(M_{D^*_{s3}}^2,M_{D}^2,q^2\right)\left( M_{D^*_{s3}}^2+M_{D}^2-q^2\right)}{40M_{D^*_{s3}}^4}\,\left(p_{\mu} g_{\nu\rho}+p_{\nu}g_{\mu\rho}+p_{\rho}g_{\mu\nu} \right) \nonumber\\
&&-\frac{\lambda\left(M_{D^*_{s3}}^2,M_{D}^2,q^2\right)}{20M_{D^*_{s3}}^2}\,\left(p^\prime_{\mu} g_{\nu\rho}+p^\prime_{\nu}g_{\mu\rho}+p^\prime_{\rho}g_{\mu\nu} \right) \nonumber\\
&&-\frac{\lambda\left(M_{D^*_{s3}}^2,M_{D}^2,q^2\right)+5M_{D^*_{s3}}^2M_{D}^2}{5M_{D^*_{s3}}^4}\,\left(p^\prime_{\mu}p_{\nu}p_{\rho}+p^\prime_{\nu}p_{\mu}p_{\rho}
+p^\prime_{\rho}p_{\mu}p_{\nu} \right) \nonumber\\
&&\left.+\frac{M_{D^*_{s3}}^2+M_{D}^2-q^2}{2M_{D^*_{s3}}^2}\,\left(p^\prime_{\mu}p^\prime_{\nu}p_{\rho}+p^\prime_{\nu}p^\prime_{\rho}p_{\mu}
+p^\prime_{\rho}p^\prime_{\mu}p_{\nu} \right) -p^\prime_{\mu}p^\prime_{\nu}p^\prime_{\rho}\right\}\nonumber\\
&&+\frac{f_D M_D^2 f_K M_K^2f_{D^*_{s2}}\,\,G_{D^*_{s2}DK}(q^2)}{(m_c+m_d)(m_d+m_s)\left(M_{D}^2-p^{\prime2}\right)\left(M_{K}^2-q^{2}\right)\left(M_{D^*_{s2}}^2-p^2\right)}
 \nonumber\\
&&\left\{ \frac{\lambda\left(M_{D^*_{s2}}^2,M_{D}^2,q^2\right)+6M_{D^*_{s2}}^2M_{D}^2}{2M_{D^*_{s2}}^4}\,p_{\mu}p_{\nu}p_{\rho}  \right.\nonumber\\
&&+\frac{\lambda\left(M_{D^*_{s2}}^2,M_{D}^2,q^2\right) }{12M_{D^*_{s2}}^2}\,\left(p_{\mu} g_{\nu\rho}+p_{\nu}g_{\mu\rho}+p_{\rho}g_{\mu\nu} \right) \nonumber\\
&&+\left(p^\prime_{\mu}p^\prime_{\nu}p_{\rho}+p^\prime_{\nu}p^\prime_{\rho}p_{\mu}
+p^\prime_{\rho}p^\prime_{\mu}p_{\nu} \right) \nonumber\\
&&\left.-\frac{M_{D^*_{s2}}^2+M_{D}^2-q^2}{M_{D^*_{s2}}^2}\,\left(p^\prime_{\mu}p_{\nu}p_{\rho}+p^\prime_{\nu}p_{\mu}p_{\rho}
+p^\prime_{\rho}p_{\mu}p_{\nu}
 \right)  \right\}\nonumber\\
&&+\frac{f_D M_D^2 f_K M_K^2f_{D^*_{s1}}\,\,G_{D^*_{s1}DK}(q^2)}{(m_c+m_d)(m_d+m_s)\left(M_{D}^2-p^{\prime2}\right)\left(M_{K}^2-q^{2}\right)\left(M_{D^*_{s1}}^2-p^2\right)}
 \nonumber\\
&&\left\{ \frac{3\left(M_{D^*_{s1}}^2+M_{D}^2-q^2\right)}{2M_{D_{s1}^*}^2}\,p_{\mu}p_{\nu}p_{\rho}  -\left(p^\prime_{\mu}p_{\nu}p_{\rho}+p^\prime_{\nu}p_{\mu}p_{\rho}
+p^\prime_{\rho}p_{\mu}p_{\nu} \right) \right\}\nonumber\\
&&+\frac{f_D M_D^2 f_K M_K^2f_{D^*_{s0}}\,\,G_{D^*_{s0}DK}(q^2)}{(m_c+m_d)(m_d+m_s)\left(M_{D}^2-p^{\prime2}\right)\left(M_{K}^2-q^{2}\right)\left(M_{D^*_{s0}}^2-p^2\right)}
 \,p_{\mu}p_{\nu}p_{\rho}+\cdots \, ,
\end{eqnarray}

\begin{eqnarray}
\Pi_{\sigma\mu\nu\rho}(p,p^\prime)&=&\frac{f_{D^*} M_{D^*} f_K M_K^2f_{D^*_{s3}}\,\,G_{D^*_{s3}D^* K}(q^2)}{(m_d+m_s)\left(M_{D^*}^2-p^{\prime2}\right)\left(M_{K}^2-q^{2}\right)\left(M_{D^*_{s3}}^2-p^2\right)}
 \nonumber\\
&&\left\{ \frac{\lambda\left(M_{D^*_{s3}}^2,M_{D^*}^2,q^2\right)}{60M_{D^*_{s3}}^2}\,\left(g_{\mu\nu}\varepsilon_{\sigma\rho\lambda\tau}p^\lambda p^{\prime\tau} +g_{\mu\rho}\varepsilon_{\sigma\nu\lambda\tau}p^\lambda p^{\prime\tau}+g_{\nu\rho}\varepsilon_{\sigma\mu\lambda\tau}p^\lambda p^{\prime\tau}\right) \right.\nonumber\\
&&+\frac{\lambda\left(M_{D^*_{s3}}^2,M_{D^*}^2,q^2\right)+ 5M_{D^*_{s3}}^2 M_{D^*}^2 }{15M_{D^*_{s3}}^4}\nonumber\\
&&\left(\varepsilon_{\sigma\rho\lambda\tau}p_{\mu}p_{\nu}p^\lambda p^{\prime\tau} +\varepsilon_{\sigma\nu\lambda\tau}p_{\mu}p_{\rho}p^\lambda p^{\prime\tau}+\varepsilon_{\sigma\mu\lambda\tau}p_{\nu}p_{\rho}p^\lambda p^{\prime\tau}\right) \nonumber\\
&&-\frac{M_{D^*_{s3}}^2+M_{D^*}^2-q^2}{6M_{D^*_{s3}}^2}\,\left(\varepsilon_{\sigma\rho\lambda\tau}p^\prime_{\mu}p_{\nu}p^\lambda p^{\prime\tau} +\varepsilon_{\sigma\rho\lambda\tau}p_{\mu}p^\prime_{\nu}p^\lambda p^{\prime\tau} +\varepsilon_{\sigma\nu\lambda\tau}p^\prime_{\mu}p_{\rho}p^\lambda
p^{\prime\tau}\right.\nonumber\\
&&\left.+\varepsilon_{\sigma\nu\lambda\tau}p_{\mu}p^\prime_{\rho}p^\lambda
p^{\prime\tau}+\varepsilon_{\sigma\mu\lambda\tau}p_{\nu}^{\prime}p_{\rho}p^\lambda p^{\prime\tau}+\varepsilon_{\sigma\mu\lambda\tau}p_{\nu}p^\prime_{\rho}p^\lambda p^{\prime\tau}\right) \nonumber\\
&&\left.+\frac{1 }{3}\,\left(\varepsilon_{\sigma\rho\lambda\tau}p^{\prime}_{\mu}p^{\prime}_{\nu}p^\lambda p^{\prime\tau} +\varepsilon_{\sigma\nu\lambda\tau}p^{\prime}_{\mu}p^{\prime}_{\rho}p^\lambda p^{\prime\tau}+\varepsilon_{\sigma\mu\lambda\tau}p^{\prime}_{\nu}p^{\prime}_{\rho}p^\lambda p^{\prime\tau}\right)  \right\}\nonumber\\
&&+\frac{f_{D^*} M_{D^*} f_K M_K^2f_{D^*_{s2}}\,\,G_{D^*_{s2}D^*K}(q^2)}{ (m_d+m_s)\left(M_{D^*}^2-p^{\prime2}\right)\left(M_{K}^2-q^{2}\right)\left(M_{D^*_{s2}}^2-p^2\right)}
 \nonumber\\
&&\left\{-\frac{1}{2}\,\left(\varepsilon_{\sigma\rho\lambda\tau}p^\prime_{\mu}p_{\nu}p^\lambda p^{\prime\tau} +\varepsilon_{\sigma\rho\lambda\tau}p_{\mu}p^\prime_{\nu}p^\lambda p^{\prime\tau} +\varepsilon_{\sigma\nu\lambda\tau}p^\prime_{\mu}p_{\rho}p^\lambda
p^{\prime\tau}\right.\right.\nonumber\\
&&\left.+\varepsilon_{\sigma\nu\lambda\tau}p_{\mu}p^\prime_{\rho}p^\lambda
p^{\prime\tau}+\varepsilon_{\sigma\mu\lambda\tau}p_{\nu}^{\prime}p_{\rho}p^\lambda p^{\prime\tau}+\varepsilon_{\sigma\mu\lambda\tau}p_{\nu}p^\prime_{\rho}p^\lambda p^{\prime\tau}\right)    \nonumber\\
&&\left. +\frac{M_{D^*_{s2}}^2+M_{D^*}^2-q^2 }{2M_{D^*_{s2}}^2}\left(\varepsilon_{\sigma\rho\lambda\tau}p_{\mu}p_{\nu}p^\lambda p^{\prime\tau} +\varepsilon_{\sigma\nu\lambda\tau}p_{\mu}p_{\rho}p^\lambda p^{\prime\tau}+\varepsilon_{\sigma\mu\lambda\tau}p_{\nu}p_{\rho}p^\lambda p^{\prime\tau}\right)   \right\}\nonumber\\
&&+\frac{f_{D^*} M_{D^*} f_K M_K^2f_{D^*_{s1}}\,\,G_{D^*_{s1}D^*K}(q^2)}{(m_d+m_s)\left(M_{D^*}^2-p^{\prime2}\right)\left(M_{K}^2-q^{2}\right)\left(M_{D^*_{s1}}^2-p^2\right)}
 \nonumber\\
&&\left(\varepsilon_{\sigma\rho\lambda\tau}p_{\mu}p_{\nu}p^\lambda p^{\prime\tau} +\varepsilon_{\sigma\nu\lambda\tau}p_{\mu}p_{\rho}p^\lambda p^{\prime\tau}+\varepsilon_{\sigma\mu\lambda\tau}p_{\nu}p_{\rho}p^\lambda p^{\prime\tau}\right) +\cdots \, ,
\end{eqnarray}
where the $\cdots$ denotes the contributions come from the higher resonances and continuum states, $\lambda(a,b,c)=a^2+b^2+c^2-2ab-2bc-2ca$, the  decay constants  $f_D$, $f_{D^*}$, $f_K$ and the hadronic coupling constants $G_{D_{s3}^*DK}$, $G_{D_{s2}^*DK}$, $G_{D_{s1}^*DK}$, $G_{D_{s0}^*DK}$, $G_{D_{s3}^*D^*K}$, $G_{D_{s2}^*D^*K}$, $G_{D_{s1}^*D^*K}$ are defined by
\begin{eqnarray}
\langle 0|J_{5}(0)|D(p^\prime)\rangle&=& \frac{f_D M_D^2}{m_c+m_d}  \, , \nonumber\\
\langle 0|J_{\sigma}(0)|D^*(p^\prime)\rangle&=& f_{D^*} M_{D^*} \varepsilon_{\sigma}(p^\prime,s) \, , \nonumber\\
\langle 0|J_{K}(0)|K(q)\rangle&=& \frac{f_K M_K^2}{m_s+m_d}  \, ,
\end{eqnarray}

\begin{eqnarray}
\langle D(p^\prime)K(q)\mid D_{s3}^*(p)\rangle &=& G_{D_{s3}^*DK} \,\varepsilon_{\alpha\beta\gamma}(p,s)\,p^{\prime\alpha}\,p^{\prime\beta} \,p^{\prime\gamma}\, , \nonumber\\
\langle D(p^\prime)K(q)\mid D_{s2}^*(p)\rangle &=& G_{D_{s2}^*DK} \,\varepsilon_{\alpha\beta }(p,s)\,p^{\prime\alpha}\,p^{\prime\beta}  \, , \nonumber\\
\langle D(p^\prime)K(q)\mid D_{s1}^*(p)\rangle &=& G_{D_{s1}^*DK} \,\varepsilon_{\alpha }(p,s)\,p^{\prime\alpha} \, , \nonumber\\
\langle D(p^\prime)K(q)\mid D_{s0}^*(p)\rangle &=& G_{D_{s0}^*DK} \, ,
\end{eqnarray}

\begin{eqnarray}
\langle D^*(p^\prime)K(q)\mid D_{s3}^*(p)\rangle &=& G_{D_{s3}^*D^*K}\,\varepsilon_{\alpha\beta\lambda\tau} \,\varepsilon^{*\alpha}(p^\prime,s^\prime)\,\varepsilon^{\beta\omega\theta}(p,s)\,p^{\lambda}\,p^{\prime\tau} \,p^{\prime}_{\omega}\,p^{\prime}_{\theta}\, , \nonumber\\
\langle D^*(p^\prime)K(q)\mid D_{s2}^*(p)\rangle &=& G_{D_{s2}^*D^*K}\,\varepsilon_{\alpha\beta\lambda\tau} \,\varepsilon^{*\alpha}(p^\prime,s^\prime)\,\varepsilon^{\beta\omega}(p,s)\,p^{\lambda}\,p^{\prime\tau}\, p^{\prime}_{\omega}\, , \nonumber\\
\langle D^*(p^\prime)K(q)\mid D_{s1}^*(p)\rangle &=& G_{D_{s1}^*D^*K}\,\varepsilon_{\alpha\beta\lambda\tau} \,\varepsilon^{*\alpha}(p^\prime,s^\prime)\,\varepsilon^{\beta}(p,s)\, p^{\lambda}\,p^{\prime\tau} \, ,
\end{eqnarray}
the $\varepsilon_{\mu\nu\rho}(p,s)$, $\varepsilon_{\mu\nu}(p,s)$ and $\varepsilon_{\mu}(p,s)$ are the mesons' polarization vectors.

Now we  rewrite the  correlation functions  $\Pi_{\mu\nu\rho }(p,p^\prime)$ and $\Pi_{\sigma\mu\nu\rho }(p,p^\prime)$ at the phenomenological side into the following form,
\begin{eqnarray}
\Pi_{\mu\nu\rho }(p,p^\prime)&=&\Pi_{DK,3}(p^2,p^{\prime2})\, p^\prime_{\mu}p^\prime_{\nu}p^\prime_{\rho} +\widetilde{\Pi}_{DK,3}(p^2,p^{\prime2})\,\left(p^\prime_{\mu}g_{\nu\rho}+p^\prime_{\nu}g_{\mu\rho}
+p^\prime_{\rho}g_{\mu\nu} \right)  \nonumber\\
&&+\Pi_{DK,3/2/1/0}(p^2,p^{\prime2})\, p_{\mu}p_{\nu}p_{\rho}+\Pi_{DK,3/2}(p^2,p^{\prime2})\,\left(p_{\mu}g_{\nu\rho}+p_{\nu}g_{\mu\rho}+p_{\rho}g_{\mu\nu} \right) \nonumber\\
&&+\Pi_{DK,3/2/1}(p^2,p^{\prime2})\,\left(p^\prime_{\mu}p_{\nu}p_{\rho}+p^\prime_{\nu}p_{\mu}p_{\rho}
+p^\prime_{\rho}p_{\mu}p_{\nu} \right) \nonumber\\
&&+\Pi_{DK,3/2}(p^2,p^{\prime2})\,\left(p^\prime_{\mu}p^\prime_{\nu}p_{\rho}+p^\prime_{\nu}p^\prime_{\rho}p_{\mu}
+p^\prime_{\rho}p^\prime_{\mu}p_{\nu} \right) \, ,  \\
\Pi_{\sigma\mu\nu\rho }(p,p^\prime)&=&\Pi_{D^*K,3}(p^2,p^{\prime2})\frac{1}{3} \left(\varepsilon_{\sigma\rho\lambda\tau}p^{\prime}_{\mu}p^{\prime}_{\nu}p^\lambda p^{\prime\tau} +\varepsilon_{\sigma\nu\lambda\tau}p^{\prime}_{\mu}p^{\prime}_{\rho}p^\lambda p^{\prime\tau}+\varepsilon_{\sigma\mu\lambda\tau}p^{\prime}_{\nu}p^{\prime}_{\rho}p^\lambda p^{\prime\tau}\right) \nonumber\\
&&+\widetilde{\Pi}_{D^*K,3}(p^2,p^{\prime2})\left(g_{\mu\nu}\varepsilon_{\sigma\rho\lambda\tau}p^\lambda p^{\prime\tau} +g_{\mu\rho}\varepsilon_{\sigma\nu\lambda\tau}p^\lambda p^{\prime\tau}+g_{\nu\rho}\varepsilon_{\sigma\mu\lambda\tau}p^\lambda p^{\prime\tau}\right) \nonumber\\
&&+\Pi_{D^*K,3/2/1}(p^2,p^{\prime2})\left(\varepsilon_{\sigma\rho\lambda\tau}p_{\mu}p_{\nu}p^\lambda p^{\prime\tau} +\varepsilon_{\sigma\nu\lambda\tau}p_{\mu}p_{\rho}p^\lambda p^{\prime\tau}+\varepsilon_{\sigma\mu\lambda\tau}p_{\nu}p_{\rho}p^\lambda p^{\prime\tau}\right) \nonumber\\
&&+\Pi_{D^*K,3/2}(p^2,p^{\prime2})\left(\varepsilon_{\sigma\rho\lambda\tau}p^\prime_{\mu}p_{\nu}p^\lambda p^{\prime\tau} +\varepsilon_{\sigma\rho\lambda\tau}p_{\mu}p^\prime_{\nu}p^\lambda p^{\prime\tau} +\varepsilon_{\sigma\nu\lambda\tau}p^\prime_{\mu}p_{\rho}p^\lambda
p^{\prime\tau}\right.\nonumber\\
&&\left.+\varepsilon_{\sigma\nu\lambda\tau}p_{\mu}p^\prime_{\rho}p^\lambda
p^{\prime\tau}+\varepsilon_{\sigma\mu\lambda\tau}p_{\nu}^{\prime}p_{\rho}p^\lambda p^{\prime\tau}+\varepsilon_{\sigma\mu\lambda\tau}p_{\nu}p^\prime_{\rho}p^\lambda p^{\prime\tau}\right)    \, ,
\end{eqnarray}
so as to  isolate the components associated with the special tensor structures which only receive contributions come from the spin-3 meson $D^*_{s3}(2860)$,
where the contributions come from the higher resonances and continuum states are neglected, the subscripts $3$, $2$, $1$ and $0$ denote that there are contributions come from the $J^P=3^-$, $2^+$, $1^-$ and $0^+$ $c\bar{s}$ mesons, respectively.  From Eqs.(15-16), we can see that the components $\Pi_{DK,3}(p^2,p^{\prime2})$, $\widetilde{\Pi}_{DK,3}(p^2,p^{\prime2})$, $\Pi_{D^*K,3}(p^2,p^{\prime2})$  and $\widetilde{\Pi}_{D^*K,3}(p^2,p^{\prime2})$ only receive contributions come from the spin-3 meson $D^*_{s3}(2860)$.
The polarization vector $\varepsilon_{\mu\nu\rho}(p,s)$ satisfies the relation $g^{\mu\nu}\varepsilon_{\mu\nu\rho}(p,s)=g^{\mu\rho}\varepsilon_{\mu\nu\rho}(p,s)=g^{\nu\rho}\varepsilon_{\mu\nu\rho}(p,s)=0$. If we multiply both sides of Eq.(5) by $g^{\mu\nu}$,
we can obtain
\begin{eqnarray}
 g^{\mu\nu}\langle 0|J_{\mu\nu\rho}(0)|D_{s3}^*(p)\rangle&\neq&f_{D_{s3}^*}g^{\mu\nu}\varepsilon_{\mu\nu\rho}(p,s)=0  \, ,
 \end{eqnarray}
the equation does not survive. We have to introduce the traceless current $\overline{J}_{\mu\nu\rho}$ by taking the following replacement,
\begin{eqnarray}
J_{\mu\nu\rho}&\rightarrow&\overline{J}_{\mu\nu\rho}=J_{\mu\nu\rho}-\frac{1}{6}g_{\mu\nu} g^{\alpha\beta}J_{\alpha\beta\rho}-\frac{1}{6}g_{\mu\rho} g^{\alpha\beta}J_{\alpha\nu\beta}-\frac{1}{6}g_{\nu\rho} g^{\alpha\beta}J_{\mu\alpha\beta} \, ,
\end{eqnarray}
then the traceless current  $\overline{J}_{\mu\nu\rho}$  satisfies the relations $g^{\mu\nu}\overline{J}_{\mu\nu\rho}=g^{\mu\rho}\overline{J}_{\mu\nu\rho}=g^{\nu\rho}\overline{J}_{\mu\nu\rho}=0$, and
 \begin{eqnarray}
  \langle 0|\overline{J}_{\mu\nu\rho}(0)|D_{s3}^*(p)\rangle&=&f_{D_{s3}^*}\varepsilon_{\mu\nu\rho}(p,s)  \, .
  \end{eqnarray}
 According to Eq.(5) and Eq.(19), we can choose either the current $J_{\mu\nu\rho}(x)$ or the current  $\overline{J}_{\mu\nu\rho}(x)$ to interpolate the $D_{s3}^*(2860)$, as the components $\Pi_{DK,3}(p^2,p^{\prime2})$, $\widetilde{\Pi}_{DK,3}(p^2,p^{\prime2})$, $\Pi_{D^*K,3}(p^2,p^{\prime2})$  and $\widetilde{\Pi}_{D^*K,3}(p^2,p^{\prime2})$ at the phenomenological side are not changed.  At the QCD side, if the current $\overline{J}_{\mu\nu\rho}(x)$ is chosen, the components $\Pi_{DK,3}(p^2,p^{\prime2})$ and $\Pi_{D^*K,3}(p^2,p^{\prime2})$ are not modified, but the components  $\widetilde{\Pi}_{DK,3}(p^2,p^{\prime2})$ and $\widetilde{\Pi}_{D^*K,3}(p^2,p^{\prime2})$ are modified remarkably. In calculations, we observe that the components  $\widetilde{\Pi}_{DK,3}(p^2,p^{\prime2})$ and $\widetilde{\Pi}_{D^*K,3}(p^2,p^{\prime2})$ cannot lead to reliable QCD sum rules and they are discarded.
The pertinent tensor structures are $p^\prime_{\mu}p^\prime_{\nu}p^\prime_{\rho}$  and $\varepsilon_{\sigma\rho\lambda\tau}p^{\prime}_{\mu}p^{\prime}_{\nu}p^\lambda p^{\prime\tau} +\varepsilon_{\sigma\nu\lambda\tau}p^{\prime}_{\mu}p^{\prime}_{\rho}p^\lambda p^{\prime\tau}+\varepsilon_{\sigma\mu\lambda\tau}p^{\prime}_{\nu}p^{\prime}_{\rho}p^\lambda p^{\prime\tau}$,    we choose the two  components $\Pi_{DK,3}(p^2,p^{\prime2})$ and $\Pi_{D^*K,3}(p^2,p^{\prime2})$ to study the hadronic coupling constants $G_{D_{s3}^*DK}$ and $G_{D_{s3}^*D^*K}$, respectively.

Now, we briefly outline  the operator product
expansion for the correlation functions $\Pi_{\mu\nu\rho}(p,p^\prime)$ and $\Pi_{\sigma\mu\nu\rho}(p,p^\prime)$  in perturbative
QCD.  We contract the quark fields in the correlation functions
$\Pi_{\mu\nu\rho}(p,p^\prime)$ and $\Pi_{\sigma\mu\nu\rho}(p,p^\prime)$ with Wick theorem firstly,
\begin{eqnarray}
\Pi_{\mu\nu\rho}(p,p^\prime)&=&  \int   d^4xd^4y e^{ip^\prime \cdot x}e^{i(p-p^\prime) \cdot (y-z)}   {\rm Tr}\left\{i\gamma_{5}U_{ij}(x-y)i\gamma_{5}S_{jk}(y-z)\Gamma_{\mu\nu\rho} C_{ki}(z-x) \right\}\mid_{z=0}\, , \nonumber\\
\Pi_{\sigma\mu\nu\rho}(p,p^\prime)&=&  \int   d^4xd^4y e^{ip^\prime \cdot x}e^{i(p-p^\prime) \cdot (y-z)}   {\rm Tr}\left\{\gamma_{\sigma}U_{ij}(x-y)i\gamma_{5}S_{jk}(y-z)\Gamma_{\mu\nu\rho} C_{ki}(z-x) \right\}\mid_{z=0}\, ,\nonumber\\
\end{eqnarray}
where
\begin{eqnarray}
\Gamma_{\mu\nu\rho}&=& \gamma_\mu\stackrel{\leftrightarrow}{\frac{\partial}{\partial z^\nu}}\stackrel{\leftrightarrow}{\frac{\partial}{\partial z^\rho}} +\gamma_\nu\stackrel{\leftrightarrow}{\frac{\partial}{\partial z^\mu}}\stackrel{\leftrightarrow}{\frac{\partial}{\partial z^\rho}}+\gamma_\rho\stackrel{\leftrightarrow}{\frac{\partial}{\partial z^\mu}}\stackrel{\leftrightarrow}{\frac{\partial}{\partial z^\nu}} \, ,\\
C_{ij}(x)&=&\frac{i}{(2\pi)^4}\int d^4k e^{-ik \cdot x} \left\{
\frac{\delta_{ij}}{\!\not\!{k}-m_c}
-\frac{g_sG^n_{\alpha\beta}t^n_{ij}}{4}\frac{\sigma^{\alpha\beta}(\!\not\!{k}+m_c)+(\!\not\!{k}+m_c)
\sigma^{\alpha\beta}}{(k^2-m_c^2)^2}\right.\nonumber\\
&&\left. +\frac{ig_s^2 GG\delta_{ij}}{12}\frac{m_ck^2+m_c^2\!\not\!{k}  }{(k^2-m_c^2)^4}+\cdots\right\}\, ,
\end{eqnarray}
  $t^n=\frac{\lambda^n}{2}$, the $\lambda^n$ is the Gell-Mann matrix, the $i$, $j$, $k$ are color indexes \cite{Reinders85}. We usually choose the full light quark propagators in the coordinate space. In the present case, the quark condensates and mixed condensates have no contributions, so we can take a simple replacement $c\rightarrow d/s$ to obtain the full $d/s$ quark propagators.  We compute  all the integrals,  then obtain the QCD spectral density through dispersion relation.

The leading-order   contributions $\Pi_{\mu\nu\rho}^{0}(p,p^{\prime})$ and $\Pi_{\sigma\mu\nu\rho}^{0}(p,p^{\prime})$ can be written as
\begin{eqnarray}
\Pi_{\mu\nu\rho}^{0}(p,p^{\prime})&=&\frac{3i}{(2\pi)^4}\int d^4k \frac{ {\rm Tr}\left\{ \gamma_5\left[ \!\not\!{k}+m_d\right]\gamma_5 \left[ \!\not\!{k}+\!\not\!{p} -\!\not\!{p^{\prime}}+ m_{s}\right]\Gamma_{\mu\nu\rho}\left[ \!\not\!{k}-\!\not\!{p^{\prime}} +m_c\right]\right\}}{\left[k^2-m_d^2\right]\left[(k+p-p^{\prime})^2-m_{s}^2\right]\left[(k-p^{\prime})^2-m_c^2\right]}\, ,\nonumber\\
&=&\int ds du \frac{\rho_{\mu\nu\rho}(s,u)}{(s-p^2)(u-p^{\prime2})} \, ,
\end{eqnarray}
\begin{eqnarray}
\Pi_{\sigma\mu\nu\rho}^{0}(p,p^{\prime})&=&\frac{3}{(2\pi)^4}\int d^4k \frac{ {\rm Tr}\left\{ \gamma_\sigma\left[ \!\not\!{k}+m_d\right]\gamma_5 \left[ \!\not\!{k}+\!\not\!{p} -\!\not\!{p^{\prime}}+ m_{s}\right]\Gamma_{\mu\nu\rho}\left[ \!\not\!{k}-\!\not\!{p^{\prime}} +m_c\right]\right\}}{\left[k^2-m_d^2\right]\left[(k+p-p^{\prime})^2-m_{s}^2\right]\left[(k-p^{\prime})^2-m_c^2\right]}\, ,\nonumber\\
&=&\int ds du \frac{\rho_{\sigma\mu\nu\rho}(s,u)}{(s-p^2)(u-p^{\prime2})} \, ,
\end{eqnarray}
where
\begin{eqnarray}
\Gamma_{\mu\nu\rho}&=&-\gamma_\mu(p-2k-2p^\prime)_\nu(p-2k-2p^\prime)_\rho-\gamma_\nu(p-2k-2p^\prime)_\mu(p-2k-2p^\prime)_\rho\nonumber\\
&&-\gamma_\rho(p-2k-2p^\prime)_\mu(p-2k-2p^\prime)_\nu\, .
\end{eqnarray}
The gluon field $G_\mu(z)$ in the covariant derivative has no contributions as $G_\mu(z)=\frac{1}{2}z^\lambda G_{\lambda\mu}(0)+\cdots=0$.
We put all the quark lines on mass-shell by using the Cutkosky's rules, see Fig.1,
and  obtain the leading-order QCD  spectral densities  $\rho_{\mu\nu\rho}(s,u)$ and $\rho_{\sigma\mu\nu\rho}(s,u)$,
\begin{eqnarray}
\rho_{\mu\nu\rho}(s,u)  &=&\frac{3}{(2\pi)^3} \int d^4k \, \delta\left[k^2-m_d^2\right]\delta\left[(k+p-p^{\prime})^2-m_{s}^2\right]\delta\left[(k-p^{\prime})^2-m_c^2\right]\nonumber\\
&& {\rm Tr}\left\{ \gamma_5\left[ \!\not\!{k}+m_{d}\right]\gamma_5 \left[ \!\not\!{k}+\!\not\!{p} -\!\not\!{p^{\prime}}+ m_{s}\right]\Gamma_{\mu\nu\rho}\left[ \!\not\!{k}-\!\not\!{p^{\prime}} +m_c\right]\right\}   \, ,
\end{eqnarray}
\begin{eqnarray}
\rho_{\sigma\mu\nu\rho}(s,u)  &=&-\frac{3i}{(2\pi)^3} \int d^4k \, \delta\left[k^2-m_d^2\right]\delta\left[(k+p-p^{\prime})^2-m_{s}^2\right]\delta\left[(k-p^{\prime})^2-m_c^2\right]\nonumber\\
&& {\rm Tr}\left\{ \gamma_\sigma\left[ \!\not\!{k}+m_{d}\right]\gamma_5 \left[ \!\not\!{k}+\!\not\!{p} -\!\not\!{p^{\prime}}+ m_{s}\right]\Gamma_{\mu\nu\rho}\left[ \!\not\!{k}-\!\not\!{p^{\prime}} +m_c\right]\right\}   \, .
\end{eqnarray}
It is straightforward to compute the integrals \footnote{ We choose the four-vectors as
$p=(\sqrt{s},0)$,  $p^\prime=(p_0^{\prime},\vec{p}^{\prime})$, $k=(k_0,\vec{k})$, and obtain the following solutions
\begin{eqnarray}
k_0&=&\frac{u-q^2+m_s^2-m_c^2}{2\sqrt{s}}\, , \,\,\, |\vec{k}|=\sqrt{ \left( \frac{u-q^2+m_s^2-m_c^2}{2\sqrt{s}}\right)^2-m_d^2  } \, , \nonumber\\
p_0^\prime&=& \frac{s+u-q^2}{2\sqrt{s}}\, , \,\,\, |\vec{p}^\prime|=\frac{\sqrt{\lambda(s,u,q^2)}}{2\sqrt{s}}\, , \nonumber
\end{eqnarray}
from the three Dirac $\delta$-functions in Eq.(26) or Eq.(27). Then we obtain   $\cos\theta$
\begin{eqnarray}
\cos\theta&=& \frac{(u-q^2+m_s^2-m_c^2)(s+u-q^2)-2s(u+m_d^2-m_c^2)}{\sqrt{(u-q^2+m_s^2-m_c^2)^2-4sm_d^2}\,\sqrt{\lambda(s,u,q^2)}} \, , \nonumber
\end{eqnarray}
 from the identity
\begin{eqnarray}
(k-p^{\prime})^2-m_c^2&=&m_d^2+u-2k_0 p_0^\prime+2  |\vec{k}| |\vec{p}^\prime| \cos\theta -m_c^2=0 \, , \nonumber
\end{eqnarray}
where we have used $\vec{k} \cdot \vec{p}^\prime =|\vec{k}| |\vec{p}^\prime| \cos\theta$. If we take the approximation $m_d^2\approx m_s^2\approx0 $,  then we obtain the constraint in Eq.(30).
}, some useful identities  are given explicitly  in the appendix.   The contributions of the gluon condensates shown in Fig.2 are calculated in the same way.

Once the analytical expressions of the QCD spectral densities are obtained,  we can take quark-hadron duality below the continuum thresholds $s_0$ and $u_0$ respectively, and perform the double Borel transform  with respect to the variables
$P^2=-p^2$ and $P^{\prime2}=-p^{\prime2}$ to obtain the QCD sum rules,
\begin{eqnarray}
\Pi_{DK,3}(M_1^2,M_2^2)&=&-\frac{f_{D}M_{D}^2f_{K}M_{K}^2f_{D^*_{s3}} \,\,G_{D^*_{s3}DK}(q^2)}{(m_c+m_d)(m_{d}+m_{s})\left(M_{K}^2-q^{2}\right)}
  \exp\left( -\frac{M_{D^*_{s3}}^2}{M_1^2}-\frac{M_{D}^2}{M_2^2}\right)\nonumber\\
  &=&\int ds du \exp\left(-\frac{s}{M_1^2}-\frac{u}{M_2^2}\right)\frac{9}{4\pi^2\sqrt{\lambda(s,u,q^2)}} \,\overline{\rho}_{DK}\, ,
\end{eqnarray}

\begin{eqnarray}
\Pi_{D^*K,3}(M_1^2,M_2^2)&=&\frac{f_{D^*}M_{D^*}f_{K}M_{K}^2f_{D^*_{s3}} \,\,G_{D^*_{s3}D^*K}(q^2)}{(m_{d}+m_{s})\left(M_{K}^2-q^{2}\right)}
  \exp\left( -\frac{M_{D^*_{s3}}^2}{M_1^2}-\frac{M_{D^*}^2}{M_2^2}\right)\nonumber\\
  &=&\int ds du \exp\left(-\frac{s}{M_1^2}-\frac{u}{M_2^2}\right)\frac{9}{4\pi^2\sqrt{\lambda(s,u,q^2)}}\, \overline{\rho}_{D^*K}\, ,
\end{eqnarray}
where
\begin{eqnarray}
\int dsdu&=&\int_{m_c^2}^{s_0} ds \int_{m_c^2}^{u_0}du \mid_{-1\leq\cos\theta\leq 1} \,\, , \nonumber\\
\cos\theta&=&\frac{\left(u-q^2-m_c^2\right)\left(s+u-q^2\right)-2s\left(u-m_c^2\right)}{|u-q^2-m_c^2|\sqrt{\lambda(u,s,q^2)}} \,\, ,
\end{eqnarray}

\begin{eqnarray}
\overline{\rho}_{DK}&=&-2m_c^2+4m_dm_c+2u+2q^2+b_1\left(4m_c^2-12m_dm_c+4m_sm_c+2s-6u-6q^2 \right)\nonumber\\
&&+b_2\left(-2m_c^2+12m_dm_c-8m_sm_c-4s+6u+6q^2 \right)\nonumber\\
&&+f_3\left(-4m_dm_c+4m_sm_c+2s-2u-2q^2 \right)\nonumber\\
&&+\frac{\pi^2}{9}\langle\frac{\alpha_sGG}{\pi}\rangle \left\{ 16\frac{\partial b_1}{\partial m_B^2}+16\frac{\partial b_1}{\partial m_A^2}+16\frac{\partial b_1}{\partial m_c^2}-17\frac{\partial b_2}{\partial m_B^2}-14\frac{\partial b_2}{\partial m_A^2}-17\frac{\partial b_2}{\partial m_c^2}
\right.\nonumber\\
&& +6\frac{\partial f_3 }{\partial m_B^2}+4\frac{\partial f_3 }{\partial m_A^2}+6\frac{\partial f_3 }{\partial m_c^2}
 +\left(3u-m_c^2-2s+9q^2 \right)\frac{\partial^2 b_2}{\partial m_A^2\partial m_B^2}\nonumber\\
&&+\left(s-u-3q^2 \right)\frac{\partial^2 f_3 }{\partial m_A^2\partial m_B^2}+\left(9u-7m_c^2-2s+3q^2 \right)\frac{\partial^2 b_2}{\partial m_A^2\partial m_c^2}
\nonumber\\
&& +\left(2m_c^2+s-3u-q^2 \right)\frac{\partial^2 f_3 }{\partial m_A^2\partial m_c^2}+\left(2s+3u-5m_c^2+3q^2 \right)\frac{\partial^2 b_2}{\partial m_B^2\partial m_c^2}\nonumber\\
&&\left.+\left(2m_c^2-s-u-q^2 \right)\frac{\partial^2 f_3 }{\partial m_B^2\partial m_c^2}-m_c^2\frac{\partial^2 b_2 }{ \partial (m_c^2)^2} +m_c^2\left(s-u-q^2 \right)\frac{\partial^3 f_3 }{ \partial (m_c^2)^3}\right\}\, ,
\end{eqnarray}

\begin{eqnarray}
\overline{\rho}_{D^*K}&=&-4m_d+4\left(m_s-m_c\right)a_1-4\left(m_c-3m_d\right)b_1+4\left(2m_c-3m_d\right)b_2+8\left(m_c-m_s \right)c_2\nonumber\\
&&+4\left(m_s-m_c\right)e_3+4\left(m_d-m_c\right)f_3\nonumber\\
&&+\frac{2\pi^2}{9}m_c\langle\frac{\alpha_sGG}{\pi}\rangle \left\{ 2\frac{\partial^2 b_2}{\partial m_A^2\partial m_B^2}+2\frac{\partial^2 c_2}{\partial m_A^2\partial m_B^2}
-\frac{\partial^2 e_3}{\partial m_A^2\partial m_B^2}-\frac{\partial^2 f_3}{\partial m_A^2\partial m_B^2}
\right.\nonumber\\
&&   -2\frac{\partial^2 b_2}{\partial m_A^2\partial m_c^2}-2\frac{\partial^2 c_2}{\partial m_A^2\partial m_c^2}+ \frac{\partial^2 e_3}{\partial m_A^2\partial m_c^2}+\frac{\partial^2 f_3}{\partial m_A^2\partial m_c^2}
\nonumber\\
&&  -2\frac{\partial^2 b_2}{\partial m_B^2\partial m_c^2}-2\frac{\partial^2 c_2}{\partial m_B^2\partial m_c^2}+ \frac{\partial^2 e_3}{\partial m_B^2\partial m_c^2}+\frac{\partial^2 f_3}{\partial m_B^2\partial m_c^2}
\nonumber\\
&&\left.+2\frac{\partial^2 b_2}{ \partial (m_c^2)^2}+2\frac{\partial^2 c_2}{ \partial (m_c^2)^2}-\frac{\partial^2 e_3}{ \partial (m_c^2)^2}-\frac{\partial^2 f_3}{ \partial (m_c^2)^2}-m_c^2\frac{\partial^3 e_3}{ \partial (m_c^2)^3}-m_c^2\frac{\partial^3 f_3}{ \partial (m_c^2)^3} \right\} \, ,
\end{eqnarray}
the explicit expressions of the  coefficients  $a_1$, $b_1$,  $b_2$, $c_2$,   $e_3$, $f_3$ are given in the appendix.

\begin{figure}
 \centering
 \includegraphics[totalheight=6cm,width=8cm]{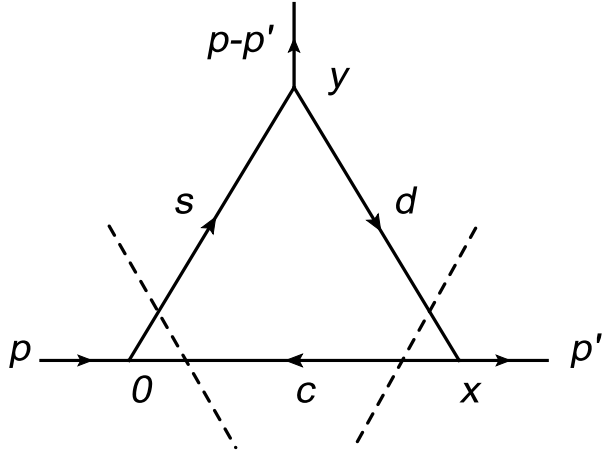}
    \caption{The leading-order   contributions, the dashed lines denote the Cutkosky's cuts.}
\end{figure}

\begin{figure}
 \centering
 \includegraphics[totalheight=8cm,width=15cm]{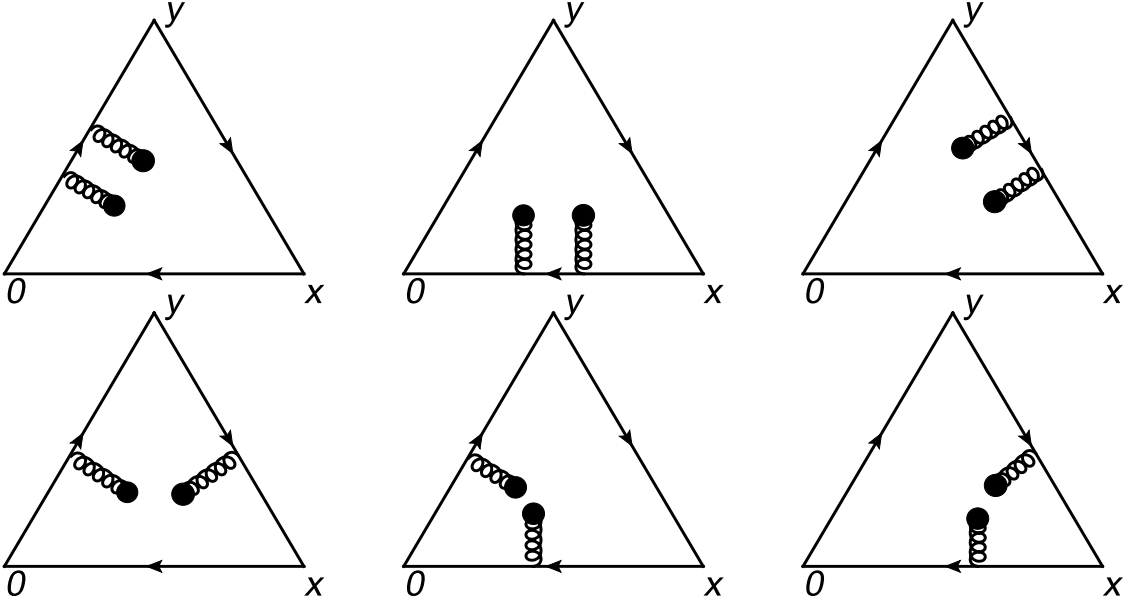}
    \caption{The gluon condensate   contributions. }
\end{figure}

\section{Numerical results and discussions}
The value  of the gluon condensate   is taken to be the standard value  $\langle \frac{\alpha_s GG}{\pi}\rangle=0.012 \,\rm{GeV}^4 $   \cite{SVZ79,Reinders85}.
 In the article, we
take the $\overline{MS}$ masses $m_{c}(m_c)=(1.275\pm0.025)\,\rm{GeV}$ and $m_s(\mu=2\,\rm{GeV})=(0.095\pm0.005)\,\rm{GeV}$
 from the Particle Data Group \cite{PDG}, and take into account
the energy-scale dependence of  the $\overline{MS}$ masses from the renormalization group equation,
\begin{eqnarray}
m_s(\mu)&=&m_s({\rm 2GeV} )\left[\frac{\alpha_{s}(\mu)}{\alpha_{s}({\rm 2GeV})}\right]^{\frac{4}{9}} \, ,\nonumber\\
m_{d}(\mu)&=&m_{d}({\rm 1GeV} )\left[\frac{\alpha_{s}(\mu)}{\alpha_{s}({\rm 1GeV})}\right]^{\frac{4}{9}} \, ,\nonumber\\
m_c(\mu)&=&m_c(m_c)\left[\frac{\alpha_{s}(\mu)}{\alpha_{s}(m_c)}\right]^{\frac{12}{25}} \, ,\nonumber\\
\alpha_s(\mu)&=&\frac{1}{b_0t}\left[1-\frac{b_1}{b_0^2}\frac{\log t}{t} +\frac{b_1^2(\log^2{t}-\log{t}-1)+b_0b_2}{b_0^4t^2}\right]\, ,
\end{eqnarray}
  where $t=\log \frac{\mu^2}{\Lambda^2}$, $b_0=\frac{33-2n_f}{12\pi}$, $b_1=\frac{153-19n_f}{24\pi^2}$, $b_2=\frac{2857-\frac{5033}{9}n_f+\frac{325}{27}n_f^2}{128\pi^3}$,  $\Lambda=213\,\rm{MeV}$, $296\,\rm{MeV}$  and  $339\,\rm{MeV}$ for the flavors  $n_f=5$, $4$ and $3$, respectively  \cite{PDG}.
  Furthermore, we obtain the values $m_u=m_d=6\,\rm{MeV}$ from the Gell-Mann-Oakes-Renner relation at the energy scale $\mu=1\,\rm{GeV}$.
  In calculations, we take $n_f=4$ and $\mu=\mu_{D_{s3}^*}=2.1\,\rm{GeV}$  \cite{Wang1603,WangEPJC-HL}.

In Ref.\cite{WangEPJC-HL}, we study the masses and decay constants of the  pseudoscalar, scalar, vector and axial-vector
heavy mesons with the QCD sum rules in a systematic way. In this article, we take the values $M_{D}=1.87\,\rm{GeV}$, $M_{D^*}=2.01\,\rm{GeV}$, $f_{D}=208\,\rm{MeV}$, $f_{D^*}=263\,\rm{MeV}$, $M^2_2(D)=(1.2-1.8)\,\rm{GeV}^2$, $M^2_2(D^*)=(1.9-2.5)\,\rm{GeV}^2$, $u_0^{D}=(6.2\pm0.5)\,\rm{GeV}^2$, $u_0^{D^*}=(6.4\pm0.5)\,\rm{GeV}^2$ determined in the two-point QCD sum rules \cite{WangEPJC-HL}.   In Ref.\cite{Wang1603}, we assign the $D_{s3}^*(2860)$ to be a D-wave $c\bar{s}$  meson, and  study the mass and decay constant (or current-meson coupling constant) of the $D_{s3}^*(2860)$ with the QCD sum rules by calculating the contributions of the vacuum condensates up to dimension-6 in the operator product expansion. In this article, we take the values  $M_{D_{s3}^*}=2.86\,\rm{GeV}$,
$f_{D_{s3}^*}=6.02\,\rm{GeV}^4$, $M^2_1(D^*_{s3})=(1.9-2.5)\,\rm{GeV}^2$, $s_0^{D^*_{s3}}=(11.6\pm0.7) \,\rm{GeV}^2$ determined in the two-point QCD sum rules \cite{Wang1603}. Furthermore, we take the values $M_K=0.495\,\rm{GeV}$ and $f_K=0.160\,\rm{GeV}$ from the Particle Data Group \cite{PDG}

\begin{figure}
 \centering
 \includegraphics[totalheight=5cm,width=6cm]{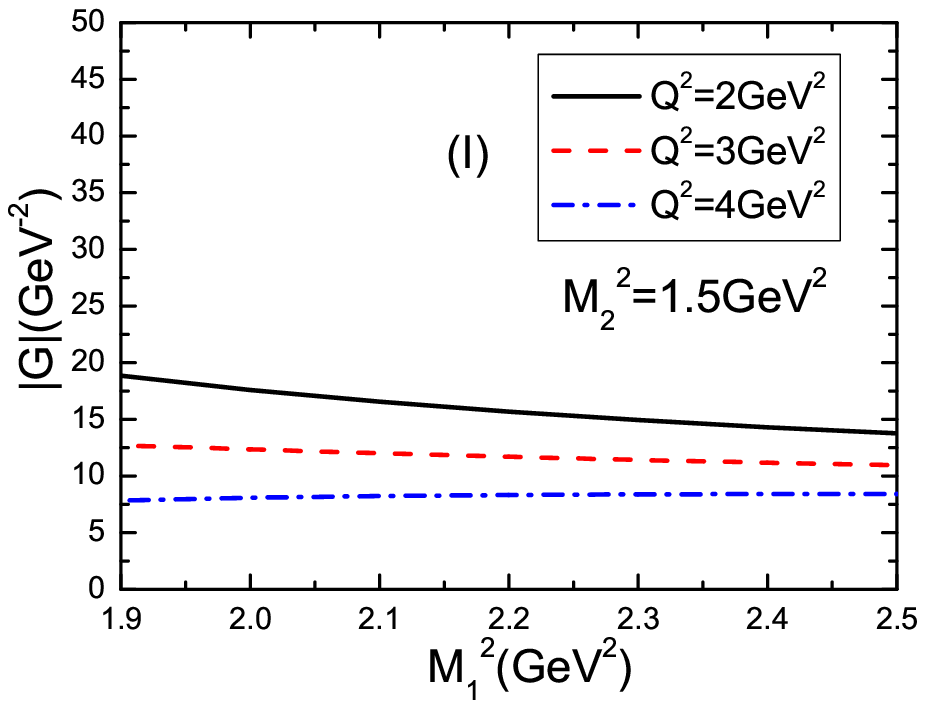}
 \includegraphics[totalheight=5cm,width=6cm]{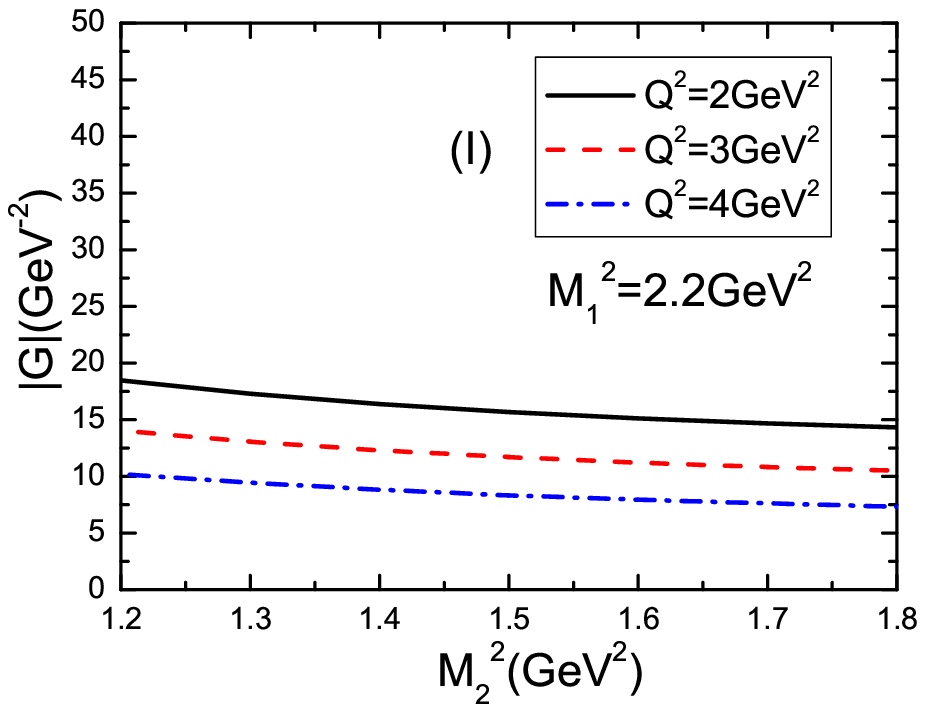}
  \includegraphics[totalheight=5cm,width=6cm]{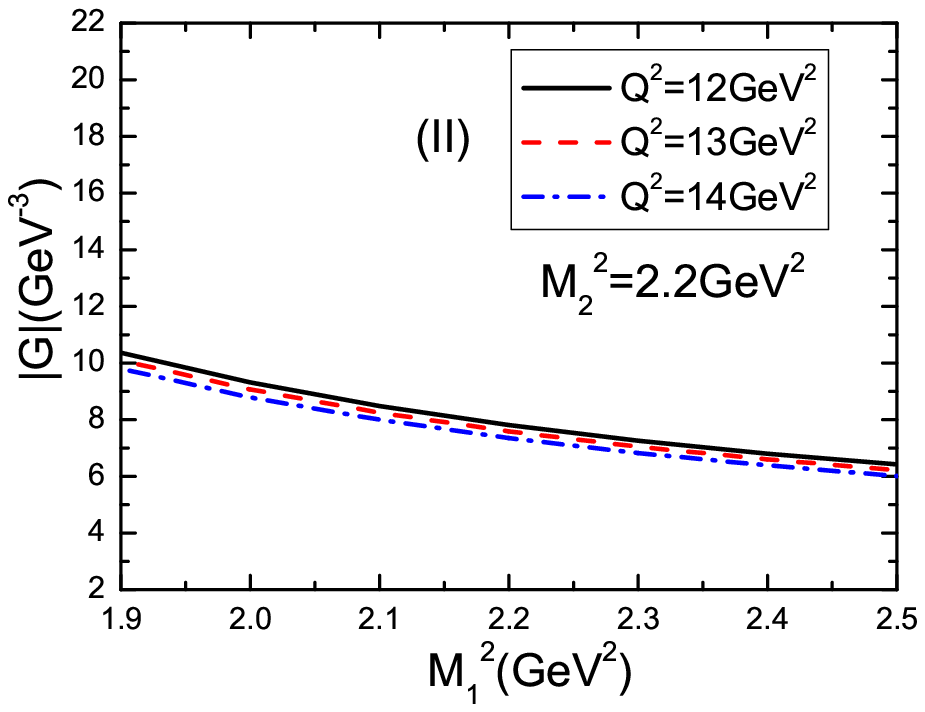}
   \includegraphics[totalheight=5cm,width=6cm]{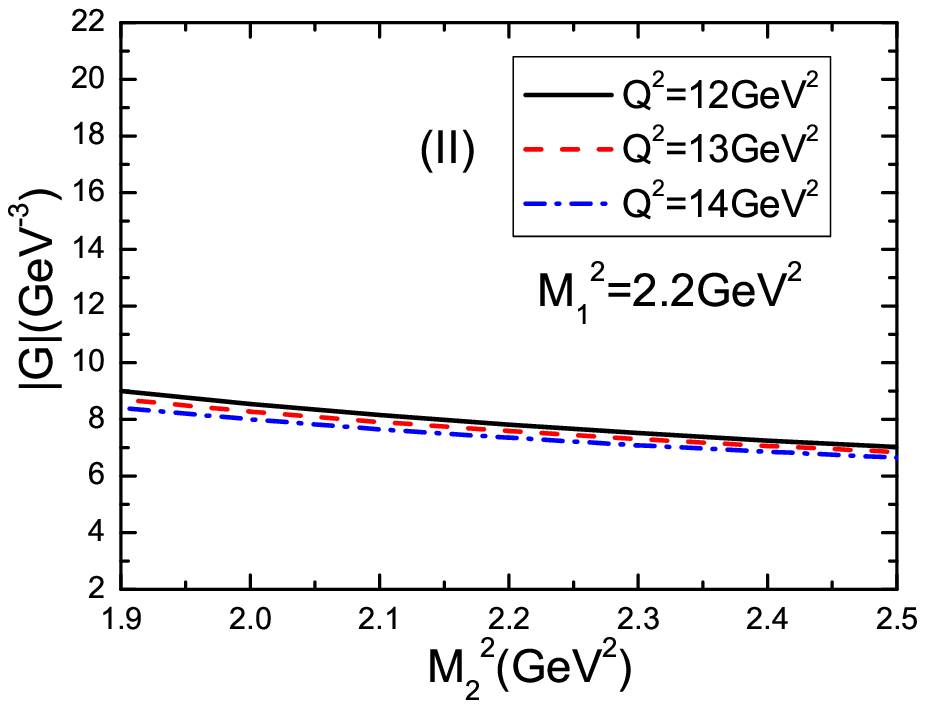}
        \caption{  The hadronic coupling constants  $G_{D^*_{s3}DK}(Q^2)$ (I) and $ G_{D^*_{s3}D^{*}K}(Q^2)$ (II) with variations of the Borel parameters $M_1^2$ and $M_2^2$, respectively.}
\end{figure}

In the following, we write down the definitions for the pole contributions of the $D_{s3}^*(2860)$, $D$ and $D^*$ in the QCD sum rules,
\begin{eqnarray}
{\rm pole}_{D_{s3}^*}&=& \frac{\int_{m_c^2}^{s_0}ds\int_{m_c^2}^{\infty}du\,\rho_{QCD}(s,u)\mid_{-1\leq\cos\theta\leq 1}\exp\left( -\frac{s}{M_1^2}-\frac{u}{M_2^2}\right)}{\int_{m_c^2}^{\infty}ds\int_{m_c^2}^{\infty}du\,\rho_{QCD}(s,u)\mid_{-1\leq\cos\theta\leq 1}\exp\left( -\frac{s}{M_1^2}-\frac{u}{M_2^2}\right)} \, , \nonumber\\
{\rm pole}_{D/D^*}&=& \frac{\int_{m_c^2}^{\infty}ds\int_{m_c^2}^{u_0}du\,\rho_{QCD}(s,u)\mid_{-1\leq\cos\theta\leq 1}\exp\left( -\frac{s}{M_1^2}-\frac{u}{M_2^2}\right)}{\int_{m_c^2}^{\infty}ds\int_{m_c^2}^{\infty}du\,\rho_{QCD}(s,u)\mid_{-1\leq\cos\theta\leq 1}\exp\left( -\frac{s}{M_1^2}-\frac{u}{M_2^2}\right)} \, ,
\end{eqnarray}
where the $\rho_{QCD}(s,u)$ denotes  the spectral densities at the QCD side. If we choose the Borel windows determined by the two-point QCD sum rules \cite{Wang1603,WangEPJC-HL}, the pole contributions  ${\rm pole}_{D_{s3}^*/D/D^*}\gg 50\%$.
For example, ${\rm pole}_{D_{s3}^*}=(92-98)\%$ for $M_2^2=1.5\,\rm{GeV}^2$ and $q^2=-3\,\rm{GeV}^2$; ${\rm pole}_{D}=(86-98)\%$ for $M_1^2=2.2\,\rm{GeV}^2$ and $q^2=-3\,\rm{GeV}^2$. The pole dominance is well satisfied.  Moreover, in the Borel windows, the contributions come from the gluon condensate are of percent level, the operator product expansion is well convergent. The Borel windows determined by the two-point QCD sum rules  still work in the three-point QCD sum rules, and we expect to make reasonable predictions.

In Fig.3, we plot the hadronic coupling constants  $G_{D^*_{s3}DK}(Q^2)$   and $ G_{D^*_{s3}D^*K}(Q^2)$   with variations of the Borel parameters $M_1^2$ and $M_2^2$, where $Q^2=-q^2$. From the figure, we can see that the values are not very stable with variations of the Borel parameters $M_1^2$ and $M_2^2$.
From the QCD sum rules in Eqs.(28-32) or  the explicit expressions of the $\overline{\rho}_{DK}$ and $\overline{\rho}_{D^*K}$, we can see that there are no contributions come from the quark condensates and mixed condensates, and no terms of the orders ${\mathcal{O}}\left(\frac{1}{M_1^2}\right)$,  ${\mathcal{O}}\left(\frac{1}{M_2^2}\right)$,  ${\mathcal{O}}\left(\frac{1}{M_1^4}\right)$,  ${\mathcal{O}}\left(\frac{1}{M_2^4}\right)$, $\cdots$, which are needed to stabilize the QCD sum rules so as to warrant a platform. The uncertainties originate from the  Borel parameters are rather large, we take them into account. In calculations, we observe that the values of the $|G_{D^*_{s3}DK}(Q^2)|$ at the region $Q^2> 1\,\rm{GeV}^2$ decrease  monotonously with increase of the $Q^2$, while the values $G_{D^*_{s3}D^*K}(Q^2)$ change sign at the region $Q^2= (1-2)\,\rm{GeV}^2$,  we have to postpone  the $Q^2$ to large values.

 Now we fit the central values of the hadronic coupling constants $G_{D^*_{s3}DK}(Q^2)$ at $Q^2=(2-4)\,\rm{GeV}^2$ and $G_{D^*_{s3}D^*K}(Q^2)$ at $Q^2=(12-14)\,\rm{GeV}^2$  into the functions of the form $A+BQ^2$,
 \begin{eqnarray}
 |G_{D^*_{s3}DK}(Q^2)|  &=&22.88\,{ \rm GeV}^{-2}-3.69\,Q^2 \,{\rm GeV}^{-4}\, , \\
 |G_{D^*_{s3}D^*K}(Q^2)|&=&10.61\,{ \rm GeV}^{-3}-0.23\,Q^2 {\rm GeV}^{-5}\, ,
 \end{eqnarray}
then we extend  the values to the physical region $Q^2=-M_K^2$, and obtain
\begin{eqnarray}
 |G_{D^*_{s3}DK}(Q^2=-M_K^2)|&=&23.8  \,{ \rm GeV}^{-2}\, , \\
 |G_{D^*_{s3}D^*K}(Q^2=-M_K^2)|&=&10.7 \,{ \rm GeV}^{-3}\, ,
 \end{eqnarray}
the uncertainties of the $G_{D^*_{s3}DK}(Q^2=-M_K^2)$ and $G_{D^*_{s3}D^*K}(Q^2=-M_K^2)$ are about $18\%$ and $28\%$, respectively.

We can take the physical values of the  hadronic coupling constants $G_{D^*_{s3}DK}$ and $G_{D^*_{s3}D^*K}$ as input parameters and study  the two-body strong decays,
which take place through relative F-wave,
\begin{eqnarray}
\Gamma\left(D_{s3}^*(2860)\to D^+K^0+D^0K^+\right)&=&\frac{1}{140\pi M_{D^*_{s3}}^2}\,G^2_{D^*_{s3}DK}\,p^7 \times 2\, , \nonumber\\
&=&28.3\pm10.2 \,\rm{MeV}\, ,\\
\Gamma\left(D_{s3}^*(2860)\to D^{*+}K^0+D^{*0}K^+\right)&=&\frac{1}{105\pi} \,G^2_{D^*_{s3}D^*K}\,p^{\prime7}\times 2\, ,\nonumber\\
&=&16.2\pm9.1\,\rm{MeV} \, ,
\end{eqnarray}
where
\begin{eqnarray}
p&=&\frac{\sqrt{\lambda\left(M_{D^*_{s3}}^2,M_{D}^2, M_{K}^2 \right)}}{2M_{D^*_{s3}}}=709\,\rm{MeV}\, ,\nonumber\\
p^\prime&=&\frac{\sqrt{\lambda\left(M_{D^*_{s3}}^2,M_{D^*}^2, M_{K}^2 \right)}}{2M_{D^*_{s3}}}=585\,\rm{MeV}\, .
\end{eqnarray}

If we saturate the decay width of the $D_{s3}^*(2860)$ with the strong decays to the final states $D^+K^0$, $D^0K^+$, $D^{*+}K^0$, $D^{*0}K^+$, the total decay width is $44.5\pm10.2 \pm 9.1 \,\rm{MeV}$, which is compatible with the width  $\Gamma_{D_{s3}^*}=(53 \pm 7 \pm 4 \pm 6)\,\rm{ MeV}$ from the LHCb collaboration \cite{LHCb7574,LHCb7712}. The predicted ratio $R$
\begin{eqnarray}
R&=&\frac{\Gamma\left(D_{s3}^*(2860)\to D^{*}K\right)}{\Gamma\left(D_{s3}^*(2860)\to DK\right)}=0.57\pm0.38\, ,
\end{eqnarray}
 which has minor overlap with the experimental value,
\begin{eqnarray}
R&=& \frac{{\rm Br}\left(D_{sJ}^*(2860)\to D^{*}K\right)}{{\rm Br}\left(D_{sJ}^*(2860)\to DK\right)}=1.10 \pm 0.15 \pm 0.19\,   ,
 \end{eqnarray}
from the BaBar  collaboration \cite{BaBar2009} due to the uncertainties, while the central value is much smaller than the experimental value.
If we assign the $D_{sJ}^*(2860)$ to be the $D_{s3}^*(2860)$, the theoretical values $R$ from the leading order heavy meson effective theory \cite{Colangelo0607},
the constituent quark model with quark-meson  effective Lagrangians \cite{Zhong2010},
the ${}^3{\rm P}_0$ model \cite{Zhang2007,Li0911,Song2014,Segovia1502,Chen1507} and
the pseudoscalar emission decay model  \cite{Godfrey2013}    are much smaller than the experimental value, see Table 1.
If we take into account the chiral symmetry breaking corrections, the experimental value can be reproduced with suitable parameters in heavy meson  effective theory     \cite{WangEPJC2860}. In the present work, we cannot reproduce  the experimental value $R=1.10 \pm 0.15 \pm 0.19$ based on the QCD sum rules, and fail to obtain additional support for assigning the $D^*_{sJ}(2860)$ to be the $D^*_{s3}(2860)$.

We have two choices to reproduce the experimental value $R=1.10 \pm 0.15 \pm 0.19$, one choice is taking into account the chiral symmetry breaking corrections by fitting the  revelent  parameters  in the heavy meson effective Lagrangians \cite{WangEPJC2860}; the other choice is introducing some $D_{s2}(2860)$ and $D_{s2}^{\prime}(2860)$ components in the $D_{sJ}^*(2860)$ beyond the $D_{s3}^*(2860)$ and the $D_{s1}^*(2860)$.
  The $J^{P}=2^{-}$ mesons $D_{s2}(2860)$ and $D_{s2}^{\prime}(2860)$   decay only to the final states $D^*K$.
 If  the $D_{sJ}^*(2860)$ consists of  at least four resonances $D_{s1}^*(2860)$, $D_{s2}(2860)$, $D_{s2}^{\prime}(2860)$,  $D_{s3}^*(2860)$, the large ratio $R=1.10 \pm 0.15 \pm 0.19$ is easy to account for, as the components $D_{s2}(2860)$ and $D_{s2}^{\prime}(2860)$ can enhance the branching fraction ${\rm Br} \left(D_{sJ}^*(2860)\to D^*K\right)$ efficaciously.

\section{Conclusion}
In this article, we assign the $D_{s3}^*(2860)$ to be a D-wave $c\bar{s}$  meson, study  the vertices $D_{s3}^*(2860)DK$ and $D_{s3}^*(2860)D^*K$ in details  to select  the pertinent  tensor structures, then calculate the hadronic coupling constants $G_{D_{s3}^*(2860)DK}$ and $G_{D_{s3}^*(2860)D^*K}$  with the three-point QCD sum rules. Finally we obtain the partial decay widths $\Gamma\left(D_{s3}^*(2860)\to D^{*}K\right)$ and $\Gamma\left(D_{s3}^*(2860)\to DK\right)$, and   the ratio $R=\Gamma\left(D_{s3}^*(2860)\to D^{*}K\right)/\Gamma\left(D_{s3}^*(2860)\to DK\right)=0.57\pm0.38$. The predicted ratio $R=0.57\pm0.38$ cannot reproduce the experimental value $R=1.10 \pm 0.15 \pm 0.19$, although the theoretical and experimental values overlap slightly with each other due to the uncertainties. Some components $D_{s2}(2860)$ and $D_{s2}^{\prime}(2860)$ are needed to reproduce the experimental value, if one would like not to resort to the chiral symmetry breaking corrections to  dispel the discrepancy.

\section*{Appendix}
The explicit expressions of the  coefficients $a_1$, $b_1$, $a_2$, $b_2$, $c_2$, $d_2$, $a_3$, $b_3$, $c_3$, $d_3$, $e_3$, $f_3$ and
\begin{eqnarray}
\frac{\partial}{\partial m_i^2  } f &\doteq& \frac{\partial}{\partial m_i^2  } f(m_A,m_B,m_c)\mid_{m_A=0;m_B=0}\, ,\nonumber\\
\frac{\partial^2}{\partial m_i^2 \partial m_j^2 } f &\doteq& \frac{\partial^2}{\partial m_i^2 \partial m_j^2 } f(m_A,m_B,m_c)\mid_{m_A=0;m_B=0}\, , \nonumber\\
\frac{\partial^3}{\partial m_i^2 \partial m_j^2\partial m_k^2 } f &\doteq& \frac{\partial^2}{\partial m_i^2 \partial m_j^2\partial m_k^2 } f(m_A,m_B,m_c)\mid_{m_A=0;m_B=0}\, ,
\end{eqnarray}
with $f(m_A,m_B,m_c)=a_1(m_A,m_B,m_c)$, $b_1(m_A,m_B,m_c)$, $a_2(m_A,m_B,m_c)$, $b_2(m_A,m_B,m_c)$,  $\cdots$, $m_i^2,m_j^2,m_k^2=m_A^2$, $m_B^2$, $m_c^2$\,.

\begin{eqnarray}
\int d^4k\, \delta^3&=&\frac{\pi}{2\sqrt{\lambda(s,u,q^2)}} \, , \nonumber\\
\int d^4k\, \delta^3 \,k_\mu&=&\frac{\pi}{2\sqrt{\lambda(s,u,q^2)}}\left[a_1(m_A,m_B,m_c)\, p_\mu+b_1(m_A,m_B,m_c)\, p^\prime_\mu \right] \, , \nonumber\\
\int d^4k\, \delta^3 \, k_{\mu} k_\nu&=&\frac{\pi}{2\sqrt{\lambda(s,u,q^2)}}\left[a_2(m_A,m_B,m_c)\, p_{\mu}p_\nu+b_2(m_A,m_B,m_c)\, p^\prime_{\mu}p^\prime_{\nu} \right. \nonumber\\
&&\left. +c_2(m_A,m_B,m_c)\,\left(p_{\mu}p^\prime_\nu+p^\prime_{\mu}p_\nu\right)+d_2(m_A,m_B,m_c)\,g_{\mu\nu} \right]\, , \nonumber\\
\int d^4k\, \delta^3 \, k_{\mu} k_\nu k_\rho&=&\frac{\pi}{2\sqrt{\lambda(s,u,q^2)}}\left[a_3(m_A,m_B,m_c)\,p_{\mu}p_{\nu}p_{\rho}  \right.\nonumber\\
&&+b_3(m_A,m_B,m_c)\,\left(p_{\mu} g_{\nu\rho}+p_{\nu}g_{\mu\rho}+p_{\rho}g_{\mu\nu} \right)   \nonumber\\
&&+c_3(m_A,m_B,m_c)\,\left(p^\prime_{\mu} g_{\nu\rho}+p^\prime_{\nu}g_{\mu\rho}+p^\prime_{\rho}g_{\mu\nu} \right) \nonumber\\
&&+d_3(m_A,m_B,m_c)\,\left(p^\prime_{\mu}p_{\nu}p_{\rho}+p^\prime_{\nu}p_{\mu}p_{\rho}
+p^\prime_{\rho}p_{\mu}p_{\nu} \right) \nonumber\\
&& +e_3(m_A,m_B,m_c)\,\left(p^\prime_{\mu}p^\prime_{\nu}p_{\rho}+p^\prime_{\nu}p^\prime_{\rho}p_{\mu}
+p^\prime_{\rho}p^\prime_{\mu}p_{\nu} \right) \nonumber\\
&&\left.+f_3(m_A,m_B,m_c)\,p^\prime_{\mu}p^\prime_{\nu}p^\prime_{\rho} \right]\, ,
\end{eqnarray}
\begin{eqnarray}
\delta^3&=& \delta\left[k^2-m_A^2 \right]\delta\left[(k+p-p^\prime)^2-m_B^2 \right]\delta\left[(k-p^\prime)^2-m_c^2 \right] \, ,
\end{eqnarray}

\begin{eqnarray}
a_1(m_A,m_B,m_c)&=& \frac{1}{\lambda(s,u,q^2)}\left[ m_c^2(u-s+q^2)+u(s-u+q^2)-2um_B^2\right.\nonumber\\
&&\left.+m_A^2(u+s-q^2)\right]\, , \nonumber\\
b_1(m_A,m_B,m_c)&=& \frac{1}{\lambda(s,u,q^2)}\left[ m_c^2(s-u+q^2)+u(u-s-2q^2)+q^2(q^2-s)\right.\nonumber\\
&&\left.-2sm_A^2+m_B^2(u+s-q^2)\right]\, ,
\end{eqnarray}

\begin{eqnarray}
a_2(m_A,m_B,m_c)&=&\frac{1}{\lambda(s,u,q^2)}\left[ (u-m_c^2)^2-2m_A^2(u+m_c^2)\right]\nonumber\\
&&+\frac{6u}{ \lambda^2(s,u,q^2)} \left\{ q^2\left[m_c^4-(u+s-q^2)m_c^2+su \right]+m_A^2m_B^2(q^2-u-s)\right.\nonumber\\
&&-m_A^2\left[ s(u-s+q^2)+m_c^2(s-u+q^2)\right]\nonumber\\
&&\left.-m_B^2\left[ u(s-u+q^2)+m_c^2(u-s+q^2)\right]\right\}\, , \nonumber\\
b_2(m_A,m_B,m_c)&=&\frac{1}{\lambda(s,u,q^2)}\left[ (u-q^2-m_c^2)^2+2m_B^2(u-q^2-m_c^2)-4sm_A^2\right]\nonumber\\
&&+\frac{6s}{ \lambda^2(s,u,q^2)} \left\{ q^2\left[m_c^4-(u+s-q^2)m_c^2+su \right]+m_A^2m_B^2(q^2-u-s)\right.\nonumber\\
&&+m_A^2\left[ s(s-u-q^2)+m_c^2(u-s-q^2)\right]\nonumber\\
&&\left.+m_B^2\left[ u(u-s-q^2)+m_c^2(s-u-q^2)\right]\right\}\, , \nonumber\\
c_2(m_A,m_B,m_c)&=&\frac{1}{\lambda(s,u,q^2)}\left[ (u-m_c^2)(m_c^2+q^2-u)+m_B^2(m_c^2-u)\right.\nonumber\\
&&\left.+m_A^2(m_c^2-q^2-m_B^2+2s+u)\right]\nonumber\\
&&-\frac{3(u+s-q^2)}{ \lambda^2(s,u,q^2)} \left\{ q^2\left[m_c^4-(u+s-q^2)m_c^2+su \right]+m_A^2m_B^2(q^2-u-s)\right.\nonumber\\
&&-m_B^2\left[m_c^2(u-s+q^2)+ u(s-u+q^2)\right]\nonumber\\
&&\left.-m_A^2\left[ m_c^2(s-u+q^2)+ s(u-s+q^2)\right] \right\}\, , \nonumber\\
d_2(m_A,m_B,m_c)&=& \frac{1}{ 2\lambda(s,u,q^2) } \left\{ q^2\left[m_c^4-(u+s-q^2)m_c^2+su \right]+m_A^2m_B^2(q^2-u-s)\right.\nonumber\\
&&+m_A^2\left[ s(s-u-q^2)+m_c^2(u-s-q^2)\right]\nonumber\\
&&\left.+m_B^2\left[ u(u-s-q^2)+m_c^2(s-u-q^2)\right]\right\}\, ,
\end{eqnarray}

\begin{eqnarray}
a_3(0,0,m_c)&=&\frac{1}{ \lambda^3(s,u,q^2)} \left\{  (m_c^2-u)^3(u-s)^3+3(m_c^2-u)^2(u-s)(u^2+3um_c^2-3us-s m_c^2)q^2\right.\nonumber\\
&&-3(m_c^2-u)\left[m_c^4(s-3u)+6um_c^2(s-u)+u^2(3s-u) \right]q^4\nonumber\\
&&\left.+\left (m_c^6+9um_c^4+9u^2m_c^2+u^3 \right)q^6 \right\}\, , \nonumber\\
b_3(0,0,m_c)&=&\frac{1}{2 \lambda^2(s,u,q^2)} \left\{  (s-m_c^2)(m_c^2-u)^2(s-u) q^2+ (m_c^2-u)(m_c^4-2sm_c^2+2um_c^2-su)\right.\nonumber\\
&&\left.q^4+m_c^2 (m_c^2+u)q^6 \right\}\, , \nonumber\\
c_3(0,0,m_c)&=&\frac{1}{2 \lambda^2(s,u,q^2)} \left\{  (m_c^2-s)(m_c^2-u)^2(s-u) q^2+ (m_c^2-u)\left[m_c^4-(s+3u)m_c^2\right.\right.\nonumber\\
&&\left.\left.+s(s+2u)\right]q^4+\left(2m_c^4-2sm_c^2-3um_c^2+su \right)q^6+m_c^2 q^8 \right\}\, ,
\end{eqnarray}

\begin{eqnarray}
d_3(0,0,m_c)&=&\frac{1}{ \lambda^3(s,u,q^2)} \left\{  (m_c^2-u)^3(s-u)^3+(m_c^2-u)^2(s-u)\left[4u^2+m_c^2(s+5u)-7us-3s^2\right]\right.\nonumber\\
&&q^2+(m_c^2-u)\left[m_c^4(3u-5s)+m_c^2(9s^2-2us-15u^2)+u(3s^2+13us-6u^2) \right]q^4\nonumber\\
&&+\left[3m_c^6+m_c^4(6u-9s)-3um_c^2(2s+5u)+u^2(5s-4u) \right]q^6\nonumber\\
&&\left.+\left (3m_c^4+6um_c^2+u^2 \right)q^8 \right\}\, ,
\end{eqnarray}

\begin{eqnarray}
e_3(0,0,m_c)&=&\frac{1}{ \lambda^3(s,u,q^2)} \left\{  (m_c^2-u)^3(u-s)^3+(m_c^2-u)^2(u-s)\left[5u^2+m_c^2(u+5s)-5us-6s^2\right]\right.\nonumber\\
&&q^2+\left[m_c^6(3s-5u)+m_c^4(9u^2+15us-6s^2)-3m_c^2(s^3+us^2+10u^2s-2u^3) \right.\nonumber\\
&&\left.+u(3s^3+9us^2+12u^2s-10u^3)\right]q^4\nonumber\\
&&+\left[3m_c^6-3m_c^4(5u+2s)+3m_c^2(3s^2+8us+2u^2)+u(10u^2-6us-5s^2) \right]q^6\nonumber\\
&&\left.+\left [6m_c^4-9m_c^2(u+s)+u(s-5u) \right]q^8+(u+3m_c^2)q^{10} \right\}\, ,
\end{eqnarray}

\begin{eqnarray}
f_3(0,0,m_c)&=&\frac{1}{ \lambda^3(s,u,q^2)} \left\{  (m_c^2-u)^3(s-u)^3+3(m_c^2-u)^2(u-s)\left[3s^2+us-2u^2+m_c^2(u-3s)\right]\right.\nonumber\\
&&q^2+3\left[m_c^6(3s-u)+m_c^4(6u^2-13us-3s^2)+m_c^2(3s^3+9us^2+12u^2s-10u^3) \right.\nonumber\\
&&\left.+u(5u^3-2u^2s-6us^2-3s^3)\right]q^4\nonumber\\
&&+\left[m_c^6+3m_c^4(5s-4u)-3m_c^2(5s^2+6us-10u^2)+6s^2u-6su^2-20u^3-s^3 \right]q^6\nonumber\\
&&\left.+3\left [m_c^4+m_c^2(s-5u)+s^2+3su+5u^2 \right]q^8+3(m_c^2-s-2u)q^{10}+q^{12} \right\}\, ,
\end{eqnarray}

\begin{eqnarray}
\frac{\partial f_3}{\partial m_A^2}&=&\frac{6s}{ \lambda^3(s,u,q^2)} \left\{  (m_c^2-u)(s-u)^2(u+s-2m_c^2)-\left[s^3+4us^2+u^2s-4u^3+m_c^4(6s-4u)\right.\right.\nonumber\\
&&\left.+m_c^2(9u^2-5s^2-8us)\right]q^2+\left[9um_c^2-2m_c^4-3sm_c^2+s^2-6u^2-us \right]q^4\nonumber\\
&&\left.+\left(s+4u-3m_c^2 \right)q^6 -q^{8} \right\}\, , \nonumber\\
\frac{\partial f_3}{\partial m_B^2} &=&\frac{3}{ \lambda^3(s,u,q^2)} \left\{  (m_c^2-u)^2(s-u)^2(3s+u)+\left[ m_c^4(3s^2+4us-3u^2)\right. \right.\nonumber\\
&& \left.+m_c^2(2u^2s-6s^3-12us^2+8u^3)+u(6s^3+9us^2-6u^2s-5u^3)\right] q^2\nonumber\\
&&+\left[ s^3-us^2+12u^2s+10u^3  +m_c^4(3u-5s)+2m_c^2(5s^2+us-6u^2)\right]q^4\nonumber\\
&&\left.-\left[m_c^4+2m_c^2(s-4u)+3s^2+10u^2+10us \right]q^6 +(3s+5u-2m_c^2)q^{8}-q^{10} \right\}\, ,
\end{eqnarray}

\begin{eqnarray}
\frac{\partial^2 e_3}{\partial m_A^2\partial m_B^2}&=&\frac{2}{ \lambda^3(s,u,q^2)} \left\{  (u-s)\left[s^3+12us^2+15u^2s+2u^3-3m_c^2(3s^2+6us+u^2)\right]\right.\nonumber\\
&& +\left[s^3-4us^2-31u^2s-8u^3+m_c^2(9u^2+12us-15s^2) \right]q^2 \nonumber\\
&&\left.+\left[3s^2+23us+12u^2+3m_c^2(s-3u)\right]q^{4}+(3m_c^2-5s-8u)q^6+2q^{8} \right\} \, ,\nonumber\\
\frac{\partial^2 e_3}{\partial m_A^2\partial m_c^2}&=&\frac{2}{ \lambda^3(s,u,q^2)} \left\{  -(s-u)^2\left[2(s^2+4us+u^2)-3m_c^2(u+3s)\right]\right.\nonumber\\
&& +\left[9s^2m_c^2-4s^3-20us^2+4u^2s+12usm_c^2+8u^3-9u^2m_c^2 \right]q^2 \nonumber\\
&&\left.+\left[12s^2-15sm_c^2+4us-12u^2+9um_c^2\right]q^{4}+(8u-4s-3m_c^2)q^6-2q^{8} \right\} \, ,\nonumber\\
\frac{\partial^2 e_3}{\partial m_B^2\partial m_c^2}&=&\frac{6}{ \lambda^3(s,u,q^2)} \left\{  2(u-m_c^2)(s-u)^2(u+s)\right.\nonumber\\
&& +\left[s^3+3us^2+5u^2s-5u^3+2m_c^2(s^2-4us+u^2) \right]q^2 \nonumber\\
&&\left.+\left[ 2m_c^2(u+s)-3(s^2+2us-u^2)\right]q^{4}+(u+3s-2m_c^2)q^6-q^{8} \right\} \, ,
\end{eqnarray}

\begin{eqnarray}
\frac{\partial^2 f_3}{\partial m_A^2\partial m_B^2}&=&\frac{6s}{ \lambda^3(s,u,q^2)} \left\{  (s-u)\left[s^2-6sm_c^2+6us+3u^2-4um_c^2\right]\right.\nonumber\\
&& \left.+\left[s^2+8us+9u^2+2m_c^2(s-4u) \right]q^2 +\left(4m_c^2-5s-9u\right)q^{4}+3q^6  \right\} \, ,\nonumber\\
\frac{\partial^2 f_3}{\partial m_A^2\partial m_c^2}&=&\frac{6s}{ \lambda^3(s,u,q^2)} \left\{  (s-u)^2(s+3u-4m_c^2)+\left(5s^2-12sm_c^2+8us-9u^2+8um_c^2 \right)q^2\right.\nonumber\\
&& \left. +\left(9u-3s-4m_c^2\right)q^{4}-3q^6  \right\} \, ,\nonumber\\
\frac{\partial^2 f_3}{\partial m_B^2\partial m_c^2}&=&\frac{6}{ \lambda^3(s,u,q^2)} \left\{  (m_c^2-u)(s-u)^2(3s+u)\right.\nonumber\\
&&  +\left[u^2s-3s^3-6us^2+4u^3+m_c^2(3s^2+4us-3u^2) \right]q^2 \nonumber\\
&&\left.+\left(5s^2-5sm_c^2+us-6u^2+3um_c^2\right)q^{4}+\left(4u-s-m_c^2 \right)q^6-q^8  \right\} \, ,
\end{eqnarray}
here we have neglected the terms $m_A^4$ and $m_B^4$ in the $a_2$, $b_2$, $c_2$ and $d_2$ as they are irreverent in present calculations.

\section*{Acknowledgements}
This  work is supported by National Natural Science Foundation,
Grant Numbers 11375063,  and Natural Science Foundation of Hebei province, Grant Number A2014502017.

\end{document}